\pdfoutput=1
\let\oldvec\vec% Store \vec in \oldvec
\documentclass{aa}
\let\vec\oldvec% Restore \vec from \oldvec

\usepackage{graphicx}
\usepackage{amssymb,amsmath}
\usepackage{txfonts}
\usepackage{natbib}
\usepackage{hyperref}
\usepackage{rotating}
\usepackage{color, colortbl}

\renewcommand{\vec}[1]{\mathbf{#1}}

\definecolor{LightCyan}{rgb}{0.88,1,1}

\definecolor{Gray}{gray}{0.9}

\DeclareSymbolFont{mymathvariables}{OT1}{ybv}{m}{it}
\SetSymbolFont{mymathvariables}{normal}{OT1}{ybv}{m}{it}
\DeclareSymbolFontAlphabet{\mathnormal}{mymathvariables}
\DeclareMathSymbol{v}{\mathalpha}{mymathvariables}{118}

\begin{document}

\title{Stopping Frequency of Type III Solar Radio Bursts in Expanding Magnetic Flux Tubes}

   \author{Hamish A. S. Reid and Eduard P. Kontar}

   \date{}

   \institute{SUPA School of Physics and Astronomy, University of Glasgow, G12 8QQ, UK}

\abstract
{}
{Understanding the properties of type III radio bursts in the solar corona and interplanetary space is one of the best ways to remotely deduce the characteristics of solar accelerated electron beams and the solar wind plasma.  One feature of all type III bursts is the lowest frequency they reach (or stopping frequency).  This feature reflects the distance from the Sun that an electron beam can drive the observable plasma emission mechanism.  The stopping frequency has never been systematically studied before from a theoretical perspective.}
{Using numerical kinetic simulations, we explore the different parameters that dictate how far an electron beam can travel before it stops inducing a significant level of Langmuir waves, responsible for plasma radio emission.  We use the quasilinear approach to model self-consistently the resonant interaction between electrons and Langmuir waves in inhomogeneous plasma, and take into consideration the expansion of the guiding magnetic flux tube and the turbulent density of the interplanetary medium.}
{We find that the rate of radial expansion has a significant effect on the distance an electron beam travels before enhanced leves of Langmuir waves, and hence radio waves, cease.  Radial expansion of the guiding magnetic flux tube rarefies the electron stream to the extent that the density of non-thermal electrons is too low to drive Langmuir wave production.  The initial conditions of the electron beam have a significant effect, where decreasing the beam density or increasing the spectral index of injected electrons would cause higher type III stopping frequencies.  We also demonstrate how the intensity of large-scale density fluctuations increases the highest frequency that Langmuir waves can be driven by the beam and how the magnetic field geometry can be the cause of type III bursts only observed at high coronal frequencies.}
{}

% {We find the extent of the radial expansion has the most significant effect on the extent an electron beam can travel before significant Langmuir wave production ceases.  The radial expansion rarefies the electron beam to the extent that the density of high-energy electrons is too low to drive the instability.  We also find that the absolute level of density fluctuations has a significant effect on the type III stopping frequency by suppressing the Langmuir wave instability.}


\titlerunning{Stopping Frequency of Type III Radio Bursts}
\authorrunning{Reid and Kontar}

\maketitle

\section{Introduction}

Type III radio bursts observed in the meter range were initially classified \citep{WildMccready1950} as an impulsive radio frequency signal that drifted from high to low frequencies with time. They were given the name `type III' because the rate they drift in frequency with time (or drift rate) was higher than `type I' and `type II' bursts.  Type III bursts are one of the most frequently observed impulsive electromagnetic signals driven by the Sun and are produced by accelerated electrons beams travelling at near-relativistic energies through the plasma of the solar system \citep[see e.g.][for reviews]{1983SoPh...89..403G,SuzukiDulk1985,1985ARA&A..23..169D,1990SoPh..130....3M,1990SoPh..130..201M, ReidRatcliffe2014}.  Understanding which properties of accelerated electron beams and the solar wind plasma give rise to certain features in type IIIs allows us to use these transient burst properties as a diagnostic tool.

The escaping radio emission responsible for type III bursts can be broadly described as two stage process \citep[first suggested by][]{GinzburgZhelezniakov1958}.  The first stage is a two-stream instability between the electron beam and the background plasma that induces a high level of Langmuir waves.  The second stage is wave-wave interactions between Langmuir waves and either ion-sound waves for fundamental emission or oppositely propagating Langmuir waves for harmonic emission.  The current theory has been refined by many authors \citep[e.g.][]{ZheleznyakovZaitsev1970, 1970SoPh...15..202S,1972SoPh...24..444Z,1976SoPh...46..515S,1980SSRv...26....3M, 1983SoPh...89..403G,1985ARA&A..23..169D, 1987SoPh..111...89M,Kontar_etal1998,2003SoPh..212..111M,2004SoPh..222..299L} and continues to develop with the advent of numerical simulations \citep[e.g][]{1976SoPh...46..323T,MagelssenSmith1977,Grognard1985,1992SoPh..137..307R,Kontar2001b,Li_etal2008,KontarReid2009,2010SoPh..267..393T,2011PhPl...18e2903T,ReidKontar2013,Ratcliffe_etal2012,LiCairns2014,2014A&A...562A..57R}.

% \subsection{Motivation}

\begin{figure}
  \includegraphics[width=0.89\columnwidth]{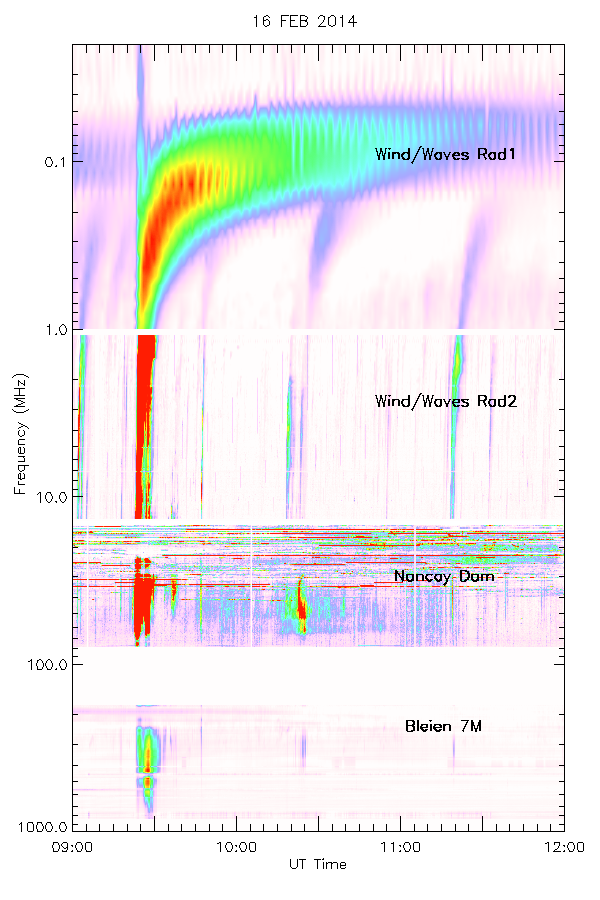}
\caption{An example of many type III bursts on the 16th February 2014, detected by the  WAVES instrument on the WIND spacecraft at 14 - 0.02 MHz \citep{Bougeret_etal1995} , the Nan\c{c}ay Decametre Array at 80 - 14 MHz \citep{Lecacheux_etal2000} and Phoenix 4 at 900 - 200 MHz \citep{Benz_etal2009}.  The large type III burst at 09:30 was associated with a GOES M1 flare.  We can observe type III bursts stopping at a variety of different frequencies.  The periodicity observed below 1 MHz is an artefact.}
\label{fig:obs1}
\end{figure}

Type III bursts cover a wide range of frequencies.  The starting frequency of type III bursts varies from hundreds of MHz, and rarely GHz, down to 1 MHz and below.  The stopping frequency of type III bursts varies over a even wider frequency range from the 100's to 10's MHz for type III bursts that originate only in the corona, all the way down to 10's kHz for type III bursts from interplanetary space.  The properties of an electron beam and the background heliospheric plasma govern at what frequencies the type III radio emission starts and stops being observed.  Whilst recent studies have investigated what is important in determining the starting frequency of type III bursts \citep{Reid_etal2011,Reid_etal2014}, the processes affecting the stopping frequency of type III bursts has never been systematically investigated.  Figure \ref{fig:obs1} shows an example of many type III bursts observed before and after a large flare.  Although electron beams associated with the flare reached frequencies around $0.04$~MHz, the majority of other type III bursts stopped at higher frequencies. What plasma parameters dictate at what frequency a type III bursts stops is unclear.

Previous observational studies on type III burst stopping frequencies by \citet{Leblanc_etal1995,Leblanc_etal1996} used radio data from the WAVES instrument onboard the WIND spacecraft and the URAP experiment on the Ulysses spacecraft \citep{Stone_etal1992}.  They analysed type III bursts detected only by Ulysses between 1990 and 1994 \citep{Leblanc_etal1995} and by both WIND and Ulysses detected below 1~MHz between 1994 and 1995 \citep{Leblanc_etal1996}.   Each type III burst was given a stopping frequency, estimated from the lowest observable frequency on the type III dynamic spectra.  Repeating the process a number of times for the same type III bursts they estimated an error on their results of 12\%.  Typical stopping frequencies for type III bursts were observed between 100-200~kHz for weak type III bursts and 20-50~kHz for strong type III bursts.  Neither study observed type III bursts below 9~kHz despite the plasma frequency at Ulysses being as low as 3~kHz during observations.  They found a connection between the intensity of the type III burst emission and the stopping frequency.  The larger the intensity of the type III burst, the lower the stopping frequency tended to be.

There were three postulations by \citet{Leblanc_etal1995} for what influences the burst stopping frequency.  In the first, the electron beam is diluted due to expansion of the guiding magnetic flux tube.  The diluted electron beam becomes ineffective in forming the required magnitude of positive slope in velocity space required for generating Langmuir waves.  In the second, the level of density fluctuations or ion-acoustic waves required to convert Langmuir waves into electromagnetic waves decreases as a function of distance from the Sun and throttles the production of radio emission (previously suggested by \citet{deGenouillacEscande1981}).  In the third, the density fluctuations suppress the level of Langmuir waves excited by the electron beam \citep[e.g.][]{1976JPSJ...41.1757N} and thus cease the production of radio emission.  That large scale density fluctuations suppress Langmuir wave growth has been discussed at length by many authors \citep[e.g.][]{SmithSime1979,Muschietti_etal1985,Kontar2001d,ReidKontar2010,Li_etal2011,Ratcliffe_etal2012}.

\citet{Dulk_etal1996} tested the hypothesis that radio wave prorogation governs the lowest type III frequencies and that the cut-offs are not intrinsic to the radiation mechanism.  Using simultaneous WIND and Ulysses (which was behind the Sun) observations, they found a reasonable correlation coefficient of 0.51 but with a high degree of scatter.  28\% of bursts they observed had the same stopping frequency to within a factor of 1.12, whilst 50\% of the bursts had the same stopping frequency to within a factor of 1.4.  At the extremes, the stopping frequency of type III bursts differed by a factor of 5.
 The difference in cutoff frequencies suggests that some directivity or rather the blocking of radio emission between source and observer by density fluctuations can play a role in determining stopping frequency.  However, it is not the dominant process because the stopping frequency of type III bursts can vary over many orders of magnitude in frequency. For example in the study by \citet{Dulk_etal1996} they recorded stopping frequencies from 300~kHz down to 20~kHz.  The large spread suggests that the plasma processes involved with the production of type III bursts largely determine the stopping frequency.

A more recent study using the WAVES instrument \citep{Bougeret_etal2008} onboard the STEREO spacecraft was carried out by \citet{Krupar_etal2014} on 154 type III bursts occurred between 2007 and 2013.  They found that  the stopping frequency in 65\% to 75\% of events was at 125~kHz, the lowest frequency channel measured.  It is also important to mention that the measured stopping frequency is dependent upon the sensitivity of the radio receiver and/or the level of ambient signal from the background plasma and the galactic background.

The necessary condition for a type III radio burst is a population of coronal accelerated electrons that are able to drive Langmuir waves with access to open magnetic field lines. If a population of accelerated electrons is able to drive Langmuir wave generation, what causes this Langmuir wave generation to cease and how does it vary with distance from the Sun?
To answer these questions, we have simulated the outward propagation of a solar electron beam injected in the low corona. We self-consistently modelled the wave-particle interactions that govern the production of Langmuir waves through the beam-plasma instability.  We also modelled collisional effects, wave refraction
in an inhomogeneous background plasma, the expanding magnetic field lines of the solar corona and solar wind.  The model and the initial condition of the injected electron beam are all described in Section \ref{sec:model_setup}.  Using Langmuir wave energy density as a proxy for the level of radio emission generated by an electron beam, we considered a number of different effects that govern when the production of Langmuir waves will cease: the expansion of guiding magnetic flux tube in Section \ref{sec:normal}, the level of density fluctuations in Section \ref{sec:inhom}, and the initial beam parameters in Section \ref{sec:beam_params}. Our results are discussed in Section \ref{sec:conclusion}.

% Is it the fundamental properties of the electron beam?  Or is it dependent upon the conditions of the background plasma?  Or a mix of the two?  To answer these questions we would ideally like to have in-situ multi-point measurements of the same electron beam throughout the inner heliosphere but this is not currently available.  Without such measurements we are left with theory to answer such questions, and the non-linear nature of the problem leans towards numerical analysis.

\section{Model equations, initial conditions and approximations}\label{sec:model_setup}

% This section will describe the model that is going to be used.  It will highlight the Langmuir wave production which is key for the radio emission.  It will not deal with the radio emission as this is too complicated right now.

\subsection{Model} \label{sec:model}

To predict the evolution of an electron beam travelling through the heliosphere we will rely on the quasilinear approximation of wave-particle interactions \citep[e.g.][]{DrummondPines1962,Vedenov_etal1962}.  This utilises the WKB approximation where we are treating waves as quasi-particles interacting resonantly $\omega_{pe}(r)=kv$, where $\omega_{pe}(r)$ is the background plasma angular frequency \citep{Bian_etal2014}.  Using $f(v,r,t)$ as the electron distribution function and $W(v,r,t)$ as the Langmuir wave spectral energy density in the one dimension of propagation, we can describe the time evolution using
\begin{eqnarray}
\frac{\partial f}{\partial t} + \frac{v}{M(r)}\frac{\partial}{\partial r}M(r)f =
\frac{4\pi ^2e^2}{m_e^2}\frac{\partial }{\partial v}\left(\frac{W}{v}\frac{\partial f}{\partial v}\right)
\quad\quad\quad\quad\quad\cr
	 +\frac{4\pi n_e e^4}{m_e^2}\ln\Lambda\frac{\partial}{\partial v}\frac{f}{v^2} + S(v,r,t)
\label{eqk1}
\end{eqnarray}
\begin{eqnarray}
\frac{\partial W}{\partial t} + \frac{\partial \omega_L}{\partial k}\frac{\partial W}{\partial r}
-\frac{\partial \omega _{pe}}{\partial r}\frac{\partial W}{\partial k}
= \frac{\pi \omega_{pe}}{n_e}v^2W\frac{\partial f}{\partial v}
 \quad\quad\quad\quad\cr
- (\gamma_{L} +\gamma_c )W + e^2\omega_{pe}v f \ln{\frac{v}{v_{Te}}}.
\label{eqk2}
\end{eqnarray}
 where $\omega_L\simeq \omega _{pe}+3k^2v_{Te}^2/(2\omega_{pe})$ is the Langmuir wave dispersion relation, $m_e$ is the electron mass, $e$ is the electron charge, $n_e(r)$ is the background plasma density (defined in  Section \ref{sec:plasma}) and $\ln\Lambda$ is the Coulomb logarithm that is assumed constant and $=20$.  $M(r)$ is the cross-sectional area of the magnetic flux tube (defined in Section \ref{sec:normal}.  In this case where $M(r)\propto r^2$ it corresponds to spherically symmetric expansion of the flux tube \citep[e.g.][]{ReidKontar2013}.  For the background electron thermal velocity we have $v_{Te}=\sqrt{k_bT_e/m_e}$ where $k_B$ is the Boltzmann constant.  The background plasma frequency is defined as $\omega_{pe}(r)=\sqrt{4\pi n_e e^2 / m_e}$.

For the particles, the physical processes we are considering in order of terms from left to right in Equation (\ref{eqk1}) are: change in time of the distribution function, transport through space in an expanding magnetic field, resonant interaction with Langmuir waves, collisional damping from the background plasma, and a source term of electrons $S(v,r,t)$  is defined in Section \ref{sec:beam}.

For the Langmuir waves, the physical processes we are considering in order of terms from left to right in Equation \ref{eqk2} are: change in time of the Langmuir wave spectral energy density, propagation of waves through space\footnote{Since Langmuir waves do not propagate far before being re-absorbed, M(r) is not taken into account in Equation (\ref{eqk2})}, Langmuir wave refraction, resonant interaction with the high velocity electrons,
Landau damping from the background Maxwellian plasma $\gamma_L=\sqrt{\pi/2}\omega_{pe}\left(v/v_{Te}\right)^3\exp\left(-{v^2}/{2v_{Te}^2}\right)$, collisional damping from background ions $\gamma_c= {\pi n_e e^4}\ln\Lambda/(m_e^2 v_{Te}^3)$, and the spontaneous generation from the background plasma.  

\subsection{Initial conditions}\label{sec:initial}

\subsubsection{Electron beam}\label{sec:beam}

We introduce a source of electrons into the simulations, where the source function takes the following form:
\begin{equation}\label{eqn:source}
S(v,r,t) = g(v)h(r)i(t).
% S(v,r,t) = Av^{-\alpha}\exp\left(\frac{(r-r_{inj})^2}{-d^2}\right)
\end{equation}

Each dimension of the source function is characterised by one parameter.  The velocity distribution is given by
\begin{equation}\label{eqn:velocity}
g(v) = A_v v^{-\alpha}\,,
\end{equation}
where $\alpha$ is the velocity spectral index.  We have chosen $\alpha$ to be 7 that corresponds to 3.5 in energy space, and is indicative of injected electron beam spectral indices derived from X-ray observations during flares \citep[e.g.][]{1988SoPh..118...49D,Holman_etal2011}.  The velocity normalisation constant is given by $A_v=n_{beam}(\alpha-1)/v_{min}^{(1-\alpha)}$.  The parameter $n_{beam}$ is the total density of electrons injected at the centre of the injection site between the minimum and maximum velocities $v_{min}=2.6v_{Te}=1.43\times10^9~\rm{cm~s}^{-1}$ and $v_{max}=36v_{Te}=2\times10^{10}~\rm{cm~s}^{-1}$.  We set $n_{beam}=2\times10^6~\rm{cm}^{-3}$.

The spatial distribution is given by
\begin{equation}\label{eqn:distance}
h(r) = \exp\left(-\frac{r^2}{d^2}\right)\,,
\end{equation}
where the characteristic parameter in space is given by the spread of the electron beam $d$.  We have chosen $d$ at $10^9$~cm which is a typical size of a flare acceleration region derived from simultaneous X-ray and radio observations \citep{Reid_etal2011,Reid_etal2014}.  The centre of the injection, $r=0$ has an altitude of $r_{inj}=50$~Mm, consistent with the estimates from a number of flares \citep{Reid_etal2011,Reid_etal2014}.  This altitude corresponds to a density of $2.14\times 10^9~\rm{cm}^{-3}$ and plasma frequency of $415$~MHz in our density model (Section \ref{sec:plasma}).  Whilst we are injecting quite a large number of electrons into the simulation ($n_{beam}/n_e=10^{-3}$), we note that the large background density will cause the majority of the electrons to lose their energy before they can escape the corona \citep{ReidKontar2013}.

The temporal profile is given by
\begin{equation}\label{eqn:time}
i(t) = A_t\exp\left(-\frac{(t-t_{inj})^2}{\tau^2}\right)
\end{equation}
where the characteristic parameter in time is given by the injection time $\tau$.  We have chosen $\tau=0.001$~s, which is equivalent to an instantaneous injection of electrons in the corona.  The time normalisation constant is given by $A_t=1/(\sqrt{\pi}\tau)$ such that $\int_{-\infty}^{\infty}i(t)dt = 1$.  An instantaneous injection was chosen to minimise the influence of injection time on the starting height that Langmuir waves are induced by the electron beam \citep{ReidKontar2013,Ratcliffe_etal2014}.  The constant $t_{inj}=4\tau$ is a delay time such that four characteristic times can occur before the injection maximum.

\subsubsection{Background plasma}\label{sec:plasma}

We introduce a population of thermal electrons as a background plasma.  This background Maxwellian population is characterised by a background temperature $T_e=2$~MK that corresponds to a background thermal velocity of $v_{Te}=5.5\times10^8~\rm{cm~s}^{-1}$.  The choice of $2$~MK is related to the higher Landau damping that is predicted from the strahl present in the heliosphere \citep[e.g.][]{Maksimovic_etal2005}.

% \textbf{EDUARD:  I want to say here that the choice of 2~MK is related to the higher Landau damping that is predicted from the strahl present in the heliosphere.  Is this how you view it?}

For the background electron density $n_e(r)$, we use the density profile of the Parker model that solves the equations for a stationary spherical symmetric solution \citep{Parker1958} with normalisation factor found from satellites \citep{Mann_etal1999}.
\begin{equation}\label{sol1}
r^2n_e(r)v(r)= C= const
\end{equation}
\begin{equation}\label{sol2}
  \frac{v(r)^2}{v_c^2}-\mbox{ln}\left(\frac{v(r)^2}{v_c^2}\right)=
  4\mbox{ln}\left(\frac{r}{r_c}\right)+4\frac{r_c}{r}-3
\end{equation}
where the critical velocity $v_c$ is defined such that $v_c\equiv (k_BT_{sw}/\tilde{\mu}m_p)^{1/2}$ and the critical radius is defined by $r_c(v_c) = GM_s/2v_c^2$ (both independent on $r$).  $T_{sw}$ is the temperature of the solar wind,\footnote{$T_{sw}$ used in the density model is different from the electron temperature $T_e$ that defines $v_{Te}$} taken as 1 MK, $M_s$ is the mass of the Sun, $m_p$ is the proton mass and $\tilde{\mu}$ is the mean molecular weight. The constant appearing above is fixed by satellite measurements near the Earth's orbit (at $r = 1$~AU, $n =6.59$~cm$^{-3}$) and equates to $6.3\times 10^{34}$~s$^{-1}$.  This model is static in time, set at the start of the simulations.  We justify this because the electron beam is moving at least two orders of magnitude faster than the solar wind velocity.

We also consider a thermal level of Langmuir waves in the background plasma \citep[e.g.][]{Hannah_etal2009,Hannah_etal2013,ReidKontar2013} when wave collisions are weak.  This takes the form
\begin{equation}\label{eqn:init_w}
W^{init}(v,r,t=0) = \frac{k_BT_e}{4\pi^2}\frac{\omega_{pe}^2}{v^2}\ln\left(\frac{v}{v_{Te}}\right),
\end{equation}
where $k_B$ is the Boltzmann constant.

\subsection{Langmuir wave energy density} \label{sec:beamplasma}

Using the model and the initial conditions described in Section \ref{sec:model} and Section \ref{sec:initial} respectively, we have simulated the propagation of an electron beam injected in the corona that can travel into interplanetary space.  The electron beam resonantly interacts with Langmuir waves forming a beam-plasma structure \citep[e.g.][]{Melnik1995,Kontar_etal1998}.  This structure consists of an electron beam inducing Langmuir waves at the front of the beam and absorbing Langmuir waves at the back of the beam.  The constant generation and absorption of Langmuir waves allows the beam-plasma structure to travel large distances up to 1~AU and beyond without running out of energy.  At any moment in time, the majority of energy is contained in the electron population and only a small fraction is contained in the Langmuir waves.

As a proxy for the type III radio bursts we use the energy density of Langmuir waves.  This is justified because only a small fraction of the energy contained in Langmuir waves is actually converted to type III radio emission.  As such, the bulk of the Langmuir wave energy is a good tracer of what times and what frequencies the electron beam will be radiating radio waves.  Moreover, dealing only with the Langmuir waves makes the problem computationally tractable for the long simulated distances of 1~AU..  It has been found that the energy density of Langmuir waves is not an exact proxy for the level of radio emission \citep{2014A&A...562A..57R,Ratcliffe_etal2014} with fundamental and harmonic type III radio emission depending differently upon the spectral characteristics of the induced Langmuir waves.  However, both branches of emission are dependent upon Langmuir waves; the absenece of enhanced Langmuir waves will stop both emission types.  The radial distance that Langmuir waves cease will give a good indication of where the beam instability terminates.  No Langmuir waves means no radio waves.

We calculated the energy density of Langmuir waves $E_w(r,t)$ (ergs cm$^{-3}$) at every point in time and space by integrating the spectral energy density $W(v,r,t)$ over $k$
\begin{equation}\label{eqn:LWenergy}
E_w(r,t) = \int_{k_{min}}^{k_{max}}{W(k,r,t)dk} = \int_{v_{min}}^{v_{max}}{\frac{\omega_{pe}}{v^2}W(v,r,t)dv}.
\end{equation}
To see the variation of Langmuir wave enhancement over thermal level as a function of distance, we normalised the level of $E_w(r,t)$ by the local thermal level of Langmuir wave energy density, found from Equation (\ref{eqn:init_w}) and (\ref{eqn:LWenergy}) as
\begin{equation}\label{eqn:LWenergy_init}
E_w^{init}(r) = \int_{v_{min}}^{v_{max}}{\frac{k_BT_e}{4\pi^2}\frac{\omega_{pe}^3}{v^4}\ln\left(\frac{v}{v_{Te}}\right)dv}.
\end{equation}
To obtain a stopping frequency for type III bursts, we must quantify at what point in space (and hence what frequency) the electron beam stops generating high levels of Langmuir waves, required to emit type III radio bursts.  We do this by calculating the maximum level of Langmuir wave energy density $E_w^{max}$ for every point of space and time.  Again we normalise this by the thermal level of Langmuir wave energy density $E_w^{init}$.  By taking a fixed level for $E_w^{max}/E_w^{init}$ we are then able to estimate the lowest frequency an electron beam was able to induce this level of Langmuir wave energy density.  We call this the stopping frequency.

\section{Radial expansion of the guiding magnetic flux tubes} \label{sec:normal}

\begin{figure*}\center
\includegraphics[width=0.49\textwidth]{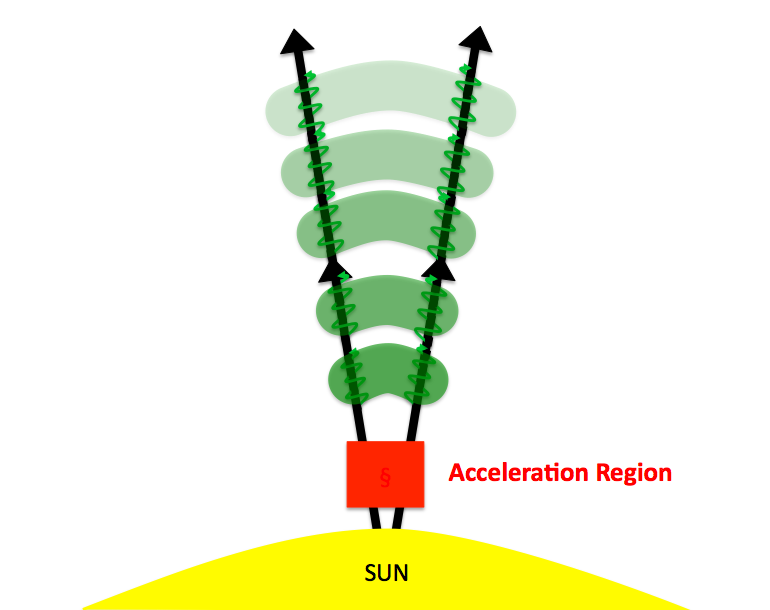}
\includegraphics[width=0.49\textwidth]{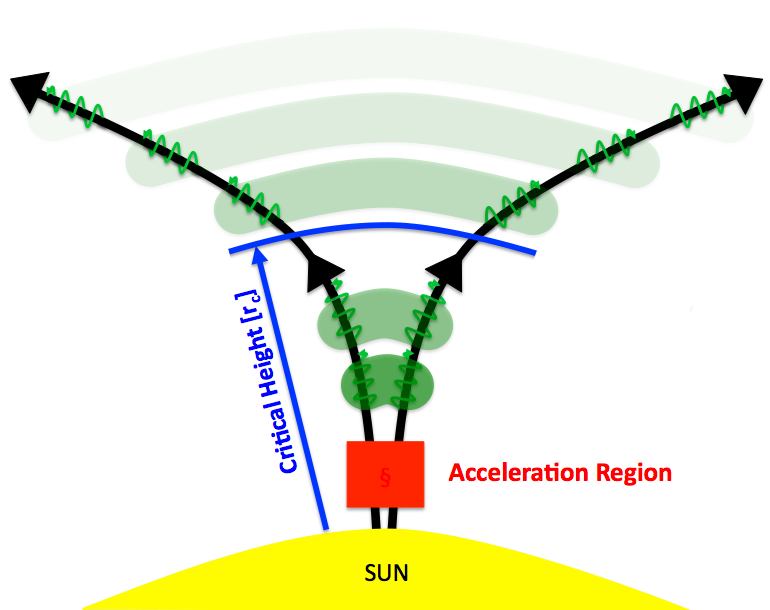}
\caption{A cartoon of an electron cloud (green) travelling upwards from an acceleration region through the corona, following the magnetic field.  The density of the electron cloud is represented through the transparency.  Left: an example of a constant radial expansion described by Equation (\ref{eqn:expansion}).  Right: At a critical height $r_c$ the magnetic field lines begin to expand at an increased rate described by Equation (\ref{eqn:double_expansion}), causing the density of the electron cloud to decrease faster.  }
\label{fig:radial_cartoon}
\end{figure*}

When electrons propagate outwards from the Sun, they follow the lines of decreasing magnetic field.  As the magnetic flux tubes expand, so too does the electron beam, causing the beam density to decrease.  An example cartoon showing this behaviour is given by Figure \ref{fig:radial_cartoon} (left).  We have modelled the expansion via the second term in Equation (\ref{eqk1}).  We use $M(r)$ to model the cross-section of the expanding flux tube that the electrons are travelling along as a function of $r$ and it takes the form
\begin{equation}\label{eqn:expansion}
M(r) = M_0\left( 1 + \frac{r}{r_0}\right)^{\beta}.
% M(r) = (r+r_0)^{\beta}.
\end{equation}
where $r_0$ is the characteristic length of the expansion of the magnetic flux tube.
The power-law index $\beta$ defines an expanding flux tube and the radial dependency of the magnetic field $B\propto M(r)^{-1} \propto r^{-\beta}$.
$M_0= \pi(d/2)^2~\rm{cm}^2 \simeq 80~\rm{Mm}^2$ is the cross-sectional area of the flux tube at the centre of the acceleration site, $r=0$.
For $r_0=3.4\times10^9$~cm, an expansion of $r^2$ gives a cone of angle 33 degrees, an angle that has been deduced from some in-situ observations of high energy electrons near the Earth \citep[e.g.][]{1974SSRv...16..189L,Steinberg_etal1985,Krucker_etal2007,Wang_etal2012}.

We have not changed the background density profile when changing the expansion of the magnetic flux tube.  Whislt the magnetic field, solar wind speed and background electron density are all interdependent properties \citep[e.g. solving equations 12-14 in][]{Kontar2001c}, it is beyond the scope of the paper to simulate how all three change with radial distance together.  We focus on the effects of the magnetic field.

As the electron cloud propagates outwards along the expanding magnetic flux
tube (Figure \ref{fig:radial_cartoon}), the electron number density decreases
with the rate dependent on the flux tube expansion rate.
Whilst velocity dispersion also causes an increases in the derivative $\partial f/\partial v$ that grows with distance,
the decrease in beam density from the electrons following the expanding magnetic flux tube is greater.
These two processes are not the entire picture but it is clear that the expansion of the magnetic flux tube
is expected to be a major factor that governs when subsequent type III radio emission stops.

%At any point in space, the growth rate of Langmuir waves is proportional to $v^2 \partial f/\partial v$.
% It is not observationally clear how the magnetic field expands as a function of distance from the Sun to the Earth.  This expansion %becomes even less clear in the corona where the structure of the magnetic field is more complex.

\subsection{Simulations}

To illustrate how the expansion of the magnetic flux tube can influence the propagation of an electron beam we numerically solved Equations (\ref{eqk1},\ref{eqk2}) with the parameters given in Section \ref{sec:model_setup} except that we change the rate of magnetic flux expansion.  For this we varied $\beta$ in Equation (\ref{eqn:expansion}) such that $\beta = 2,~2.5,~3,~3.5$.

\begin{figure*}\center
  \includegraphics[width=0.47\textwidth,trim=22 20 85 35,clip]{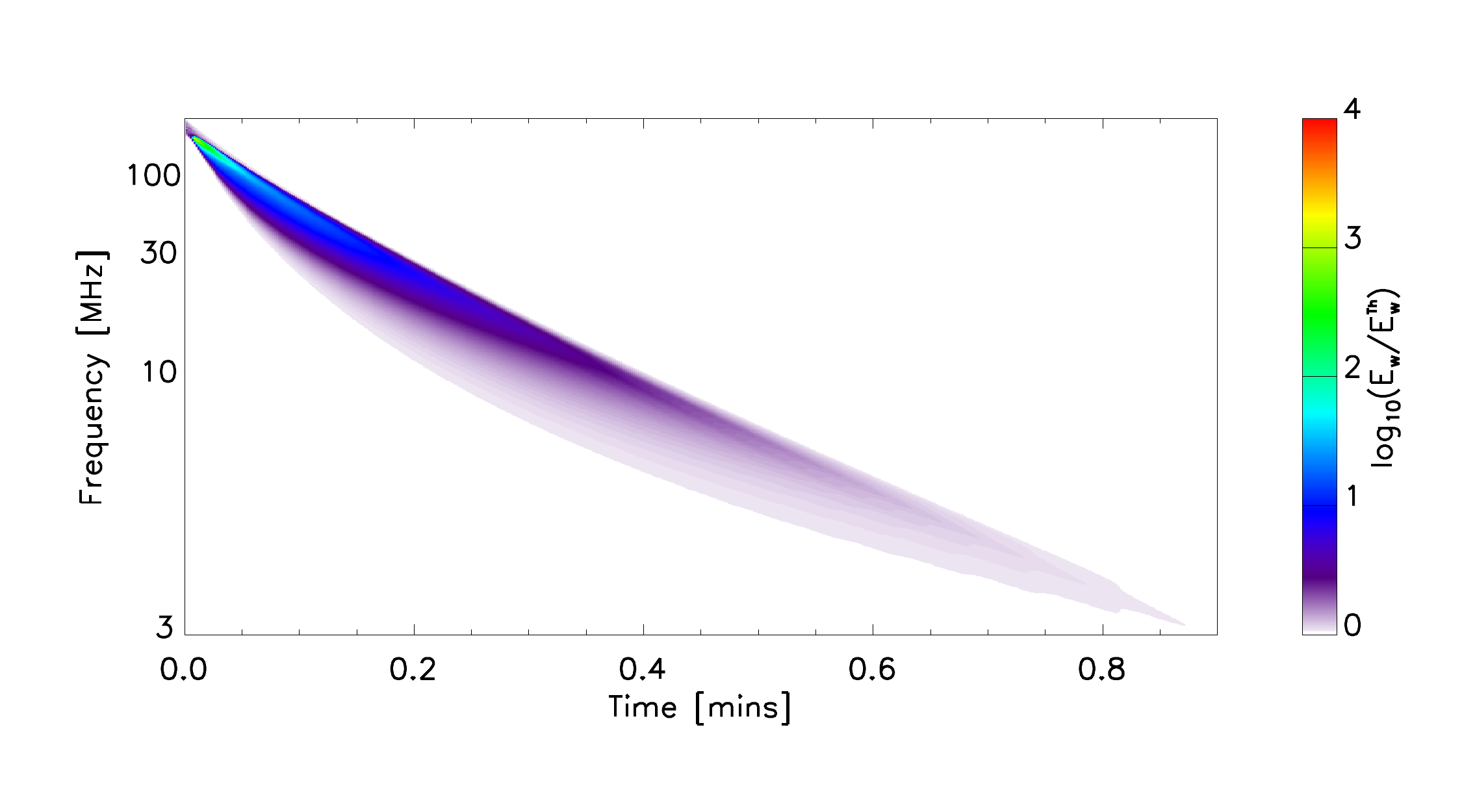}
  \includegraphics[width=0.52\textwidth,trim=22 20 25 35,clip]{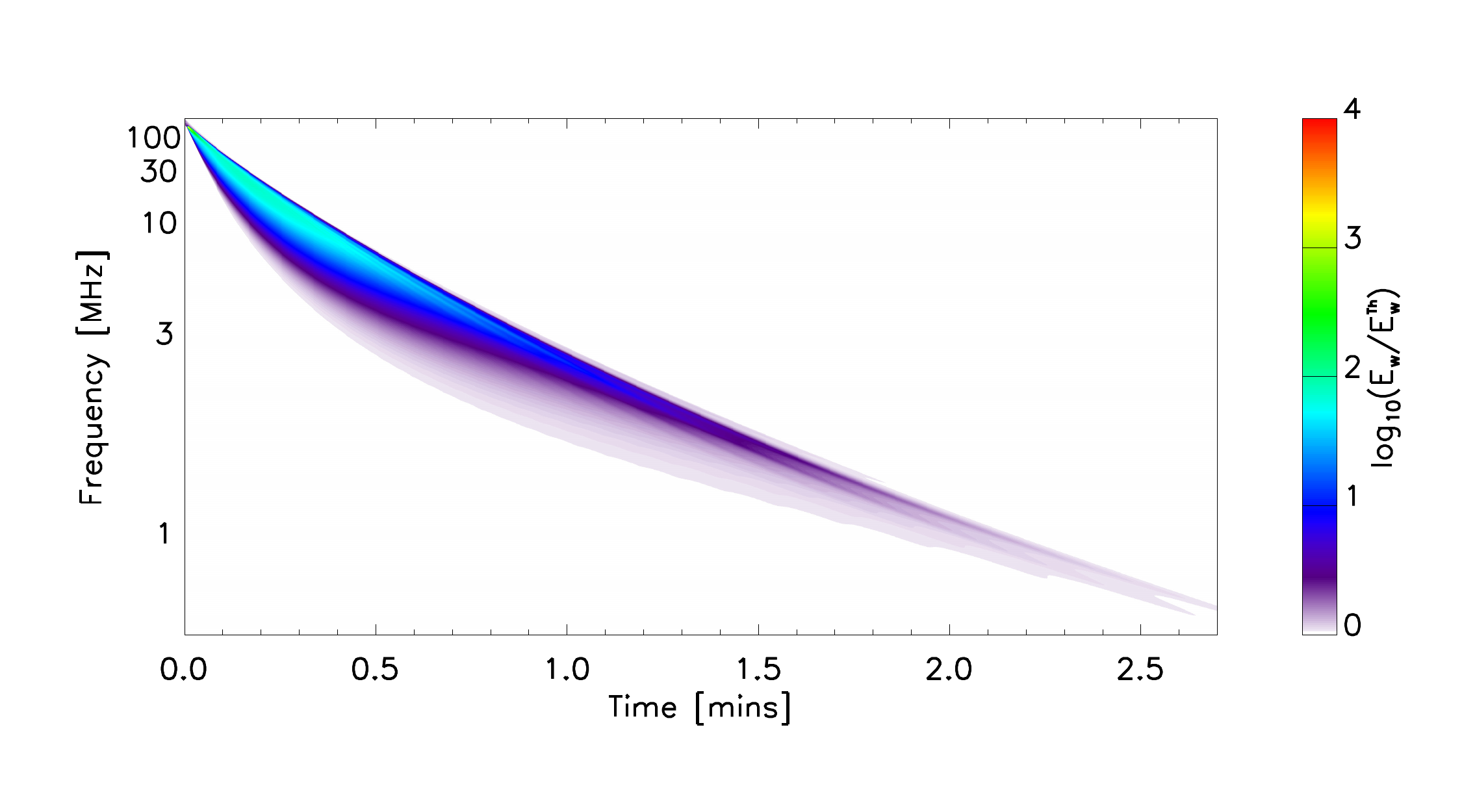}
  \includegraphics[width=0.47\textwidth,trim=22 20 85 35,clip]{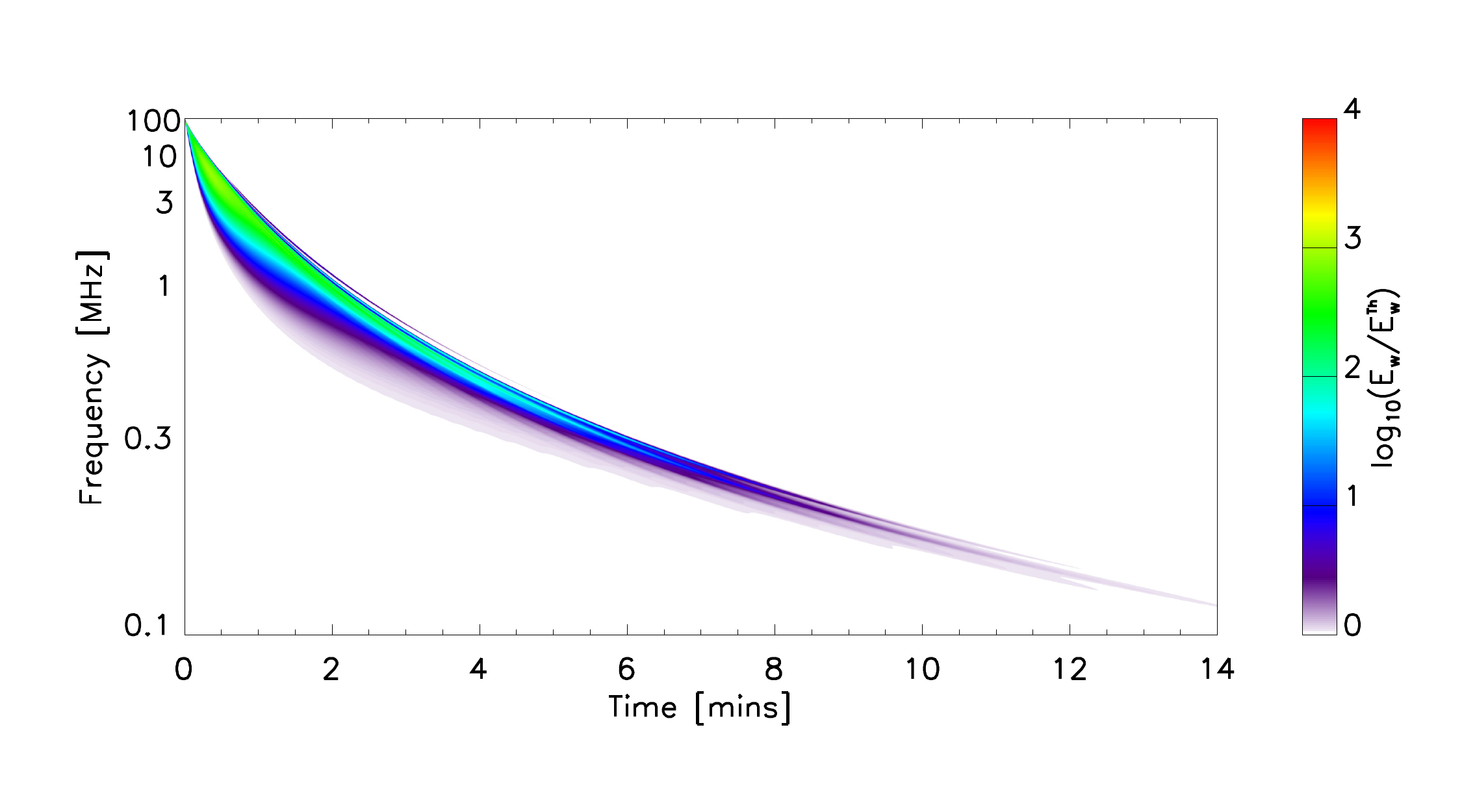}
  \includegraphics[width=0.52\textwidth,trim=22 20 25 35,clip]{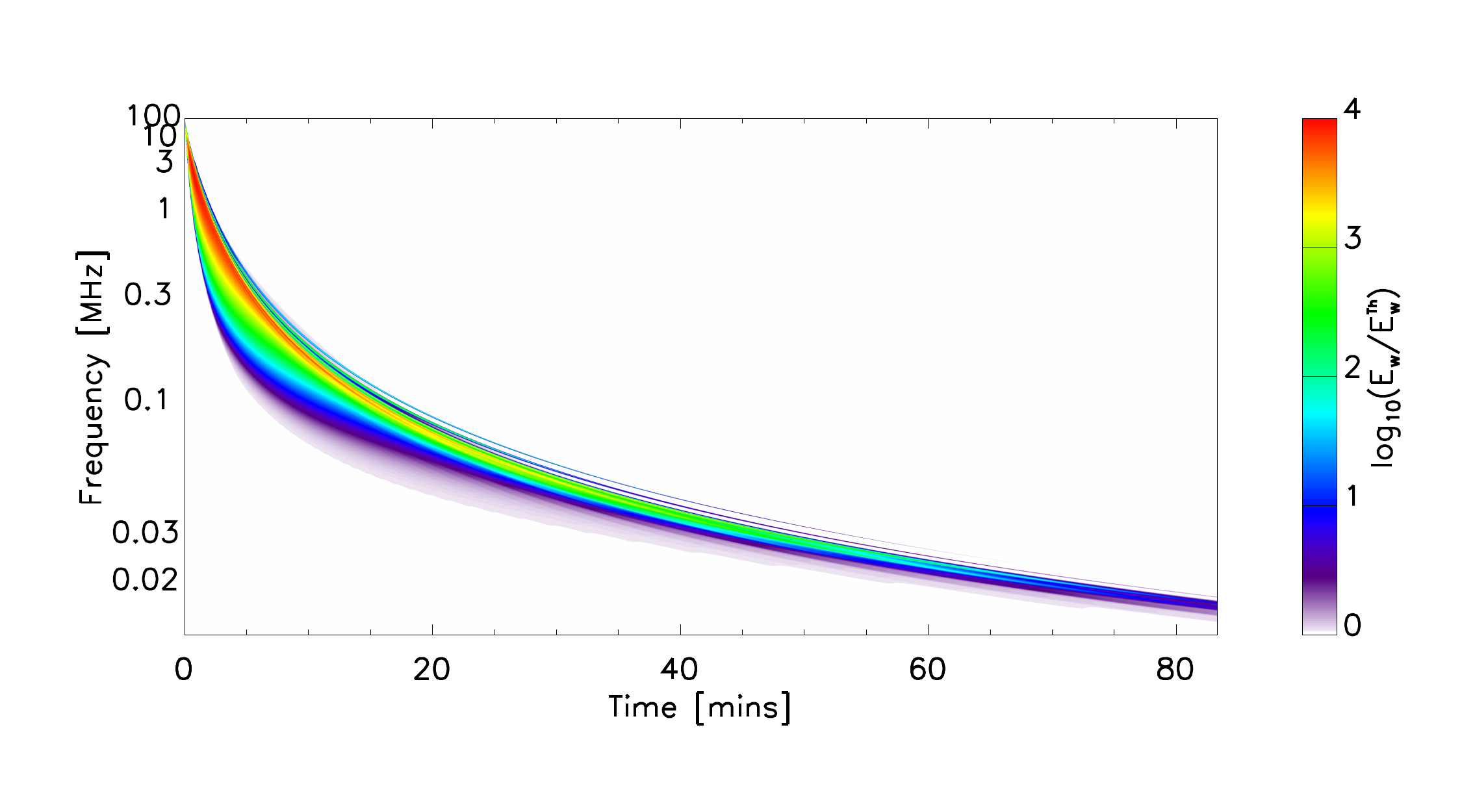}
\caption{Langmuir wave energy density $E_w /E_w^{init}$ in plasma frequency - time plane. The same electron beam parameters are used 
for all panels but with a magnetic flux tube expanding at different rates. The coefficient that models the expansion of the flux tubes is $\beta=3.5$ (top left), $\beta=3.0$ (top right), $\beta=2.5$ (bottom left), $\beta=2.0$ (bottom right).  Note the different values of the frequency and time axis between each graph.}
\label{fig:radial1}
\end{figure*}

As described in Section \ref{sec:beamplasma} we used the Langmuir wave energy density as a proxy for the type III radio emission.  We plot the Langmuir wave energy density variation as a function of time
and frequency (see e.g. Figure \ref{fig:radial1}), similar to what is normally produced for type III bursts using the flux density of radio waves as a function of time and frequency.  To account for the decreasing Langmuir wave energy as a function of frequency, shown in Equation (\ref{eqn:LWenergy}), we normalised the energy density by the thermal level of Langmuir waves given by $E_w^{init}=E_w(t=0)=\int{W_{Th}dk}$.

Figure \ref{fig:radial1} shows how the energy density of Langmuir waves varies as a function of frequency and time for all four values of $\beta = 2,~2.5,~3,~3.5$.  When $\beta=3.5$ (top left graph) the energy density in Langmuir waves is low, ceasing completely after only 0.8 mins and not reaching frequencies less than roughly 4 MHz.  The magnetic flux tube expands very rapidly, causing the electron beam to decrease in density over a short distance and is consequently no longer able to drive the wave-particle interaction.  When $\beta=3.0$ (top right graph) the magnetic flux tube expands less rapidly and Langmuir waves are induced by the electron beam over a longer time (2.5 mins) and at lower frequencies (1 MHz).  This trend continues for $\beta=2.5$ (bottom left graph) and $\beta=2.0$ (bottom right graph) with the Langmuir wave energy density being significant over the thermal level at 0.1 MHz and 0.03 MHz respectively.

\begin{figure*}
  \includegraphics[width=0.49\textwidth]{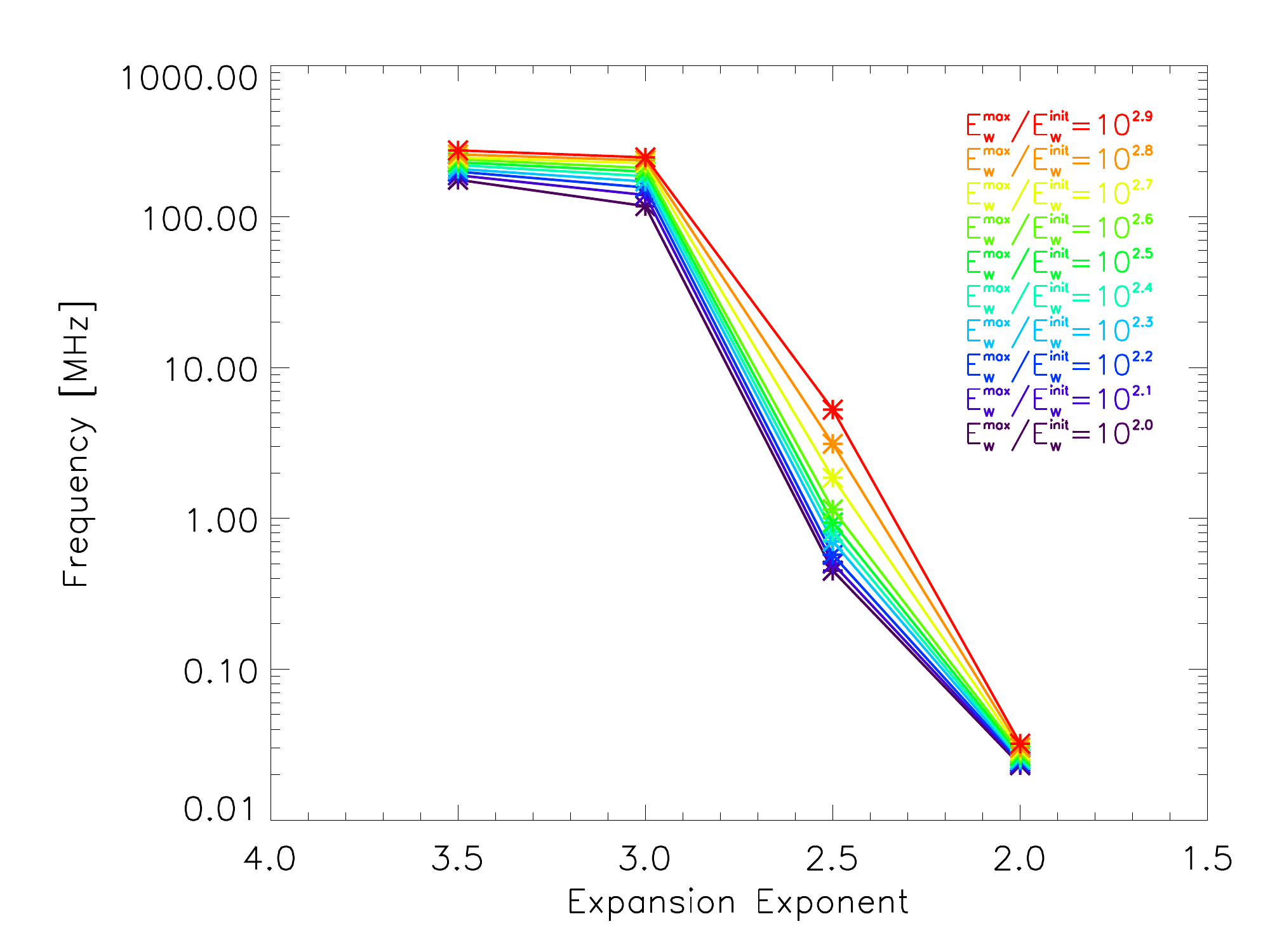}
  \includegraphics[width=0.49\textwidth]{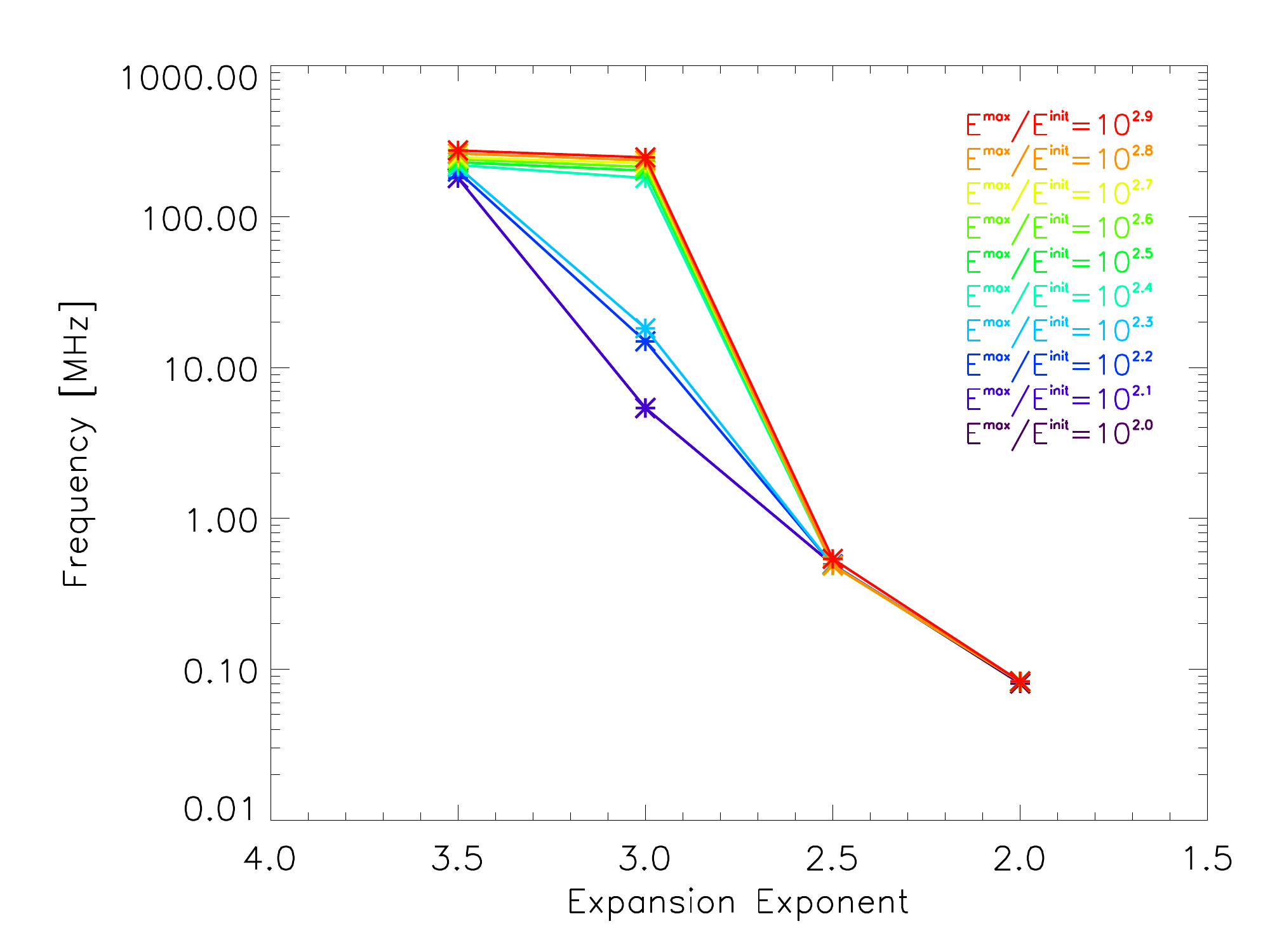}
\caption{ The minimum frequency (stopping frequency) where an electron beam is able to induce certain amount of Langmuir wave energy density (see color-coded lines). plotted against the different radial expansion exponents, $\beta$.  The different coloured lines represent different levels of Langmuir wave energy (normalised by the thermal level) between 100 and 1000.  The right graph includes density fluctuations in the background plasma.}
\label{fig:radial2}
\end{figure*}

For different levels of magnetic flux tube expansion, Figure \ref{fig:radial2} (left) shows the minimum background plasma frequency the electron beam was able to excite a certain level of Langmuir wave energy, as described in Section \ref{sec:beam}.  We note that the background plasma frequency relates to distance using the density model described in Section \ref{sec:plasma}.  The different lines in Figure \ref{fig:radial2} illustrate different values of $E_w^{max}/E_w^{init}$,  taken between $10^2$ and $10^3$.

We can see from Figure \ref{fig:radial2} that slow expansion (smaller $\beta$) leads to a significant energy of Langmuir waves at frequencies below $\sim 1$~MHz.  The points representing different levels of maximum Langmuir wave energy density $E_w^{max}/E_w^{init}$ are closer together in minimum frequency for $\beta=2$ compared to $\beta=2.5$.  As the electron beam propagates farther from the Sun, the plasma density profile becomes flatter,
($|\partial \omega_{pe}/\partial r|$ decreases) and hence the Langmuir wave energy decreases faster with frequency.  The similarity between the simulations with $\beta=3.5$ and $\beta=3.0$ is because the electron beam density decreases very quickly in both simulations.  In comparison to the the cases $\beta=2,~2.5$, only a small amount of energy is transferred to Langmuir waves before the wave generation is quenched.

When we consider weaker radial expansion (for instance $\beta=2$), we observe Langmuir waves being generated at distances much farther from the Sun and consequently at much smaller frequencies.    We considered radial expansion with smaller $\beta$ values of $\beta=1.5,~1.0$ but they produced so much Langmuir wave energy that by 2~AU (the end of our simulation box) there was still a significant level of energy density in Langmuir waves, with $E_w^{max}/E_w^{init} > 10^3$.  As shown in the next section, part of this result is related to not considering the density fluctuations in the background plasma.  These results support  the idea
 \citep{Buttighoffer_etal1995,Buttighoffer1998}
 that narrow flux tubes are encountered when electrons producing type III emission are observed with Ulysses at distances greater than 1~AU.

% Perhaps there should be a little bit more information written about the above fact..............

\subsection{Complex Radial Expansion} \label{sec:expansion}

In Equation (\ref{eqn:expansion}) the magnetic flux tube expands in the corona at the same rate as it expands in the heliosphere.  Whilst this might be more relevant in areas of open field (for instance the coronal holes), solar active regions can be far from simplistic.  It is possible that the magnetic field can be focussed in the corona through narrow flux tubes.  From Section \ref{sec:normal} we know this will facilitate a higher level of Langmuir wave turbulence. However, it has been observed that type III bursts can occur at high frequencies and not at lower frequencies.  One explanation for this behaviour is an increased expansion as electrons propagate out of the corona. Figure \ref{fig:radial_cartoon} (right) shows a cartoon of the geometry of the magnetic field in such a scenario, where the expansion of the magnetic field lines increase at a certain height $r_c$.

We model the magnetic field proposed in Figure \ref{fig:radial_cartoon} (right) by modifying magnetic flux tube cross-sectional area $M(r)$, so that
\begin{equation}\label{eqn:double_expansion}
M(r) = M_0 \begin{cases} \left( 1 + \frac{r}{r_0}\right)^{\beta} & \mbox{for } r<r_c \\ \left[ \left(1 + \frac{r}{r_0}\right)^{\beta} + Z_c \left( 1 - \frac{r}{r_c}\right)^{\beta}\right] & \mbox{for } r\geq r_c. \end{cases}
\end{equation}
where $r_c$ is a height above the photosphere where the expansion of the magnetic flux tube changes.  Above this height the expansion is more rapid, with the severity of this increased expansion modelled through
the constant $Z_c$. Similar to Equation (\ref{eqn:expansion}), the constant $M_0=80~\rm{Mm}$ is the cross-sectional area of the flux tube at the centre of the acceleration site $r=0$.

\begin{figure*}\center
  \includegraphics[width=0.47\textwidth,trim=22 20 85 35,clip]{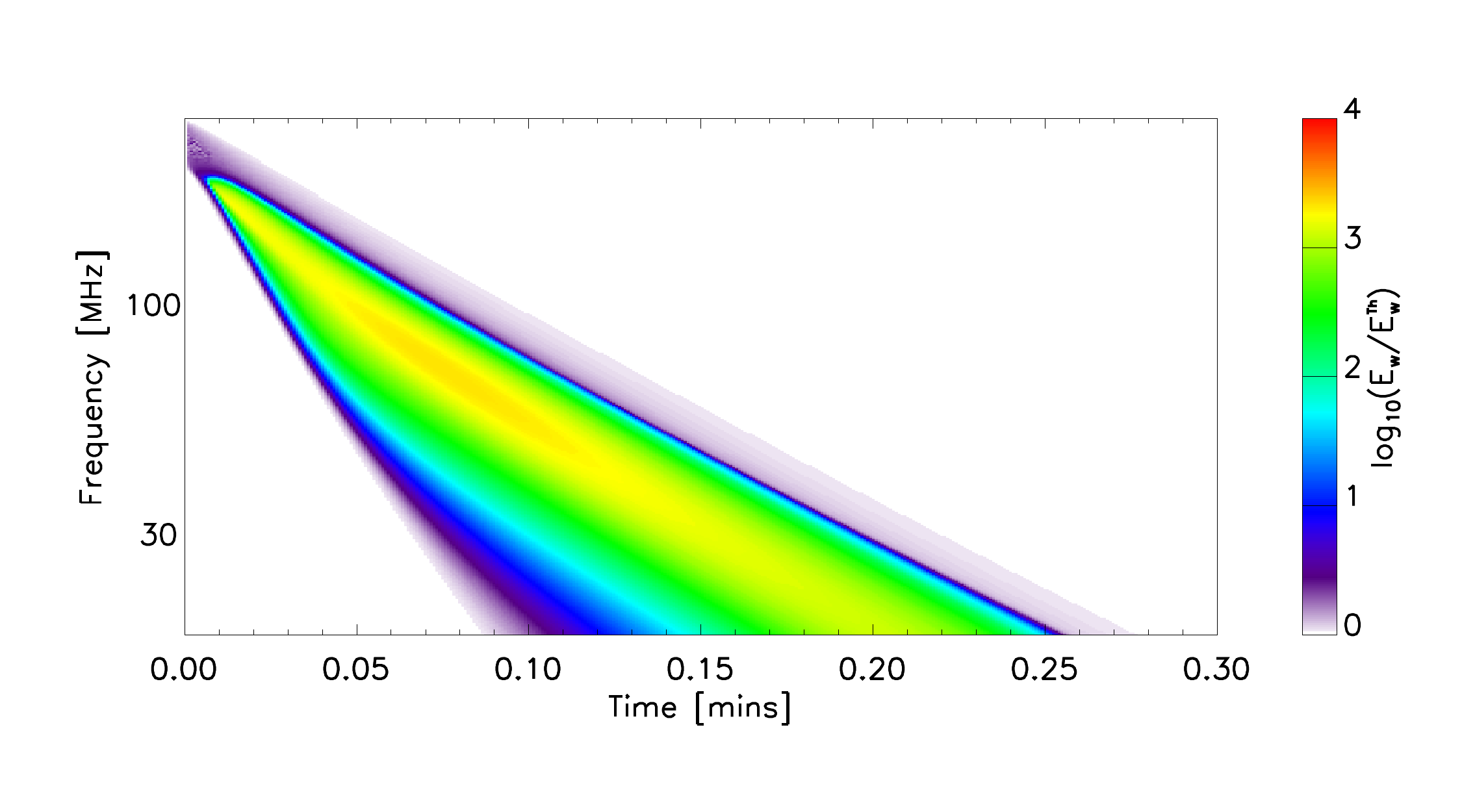}
  \includegraphics[width=0.52\textwidth,trim=22 20 25 35,clip]{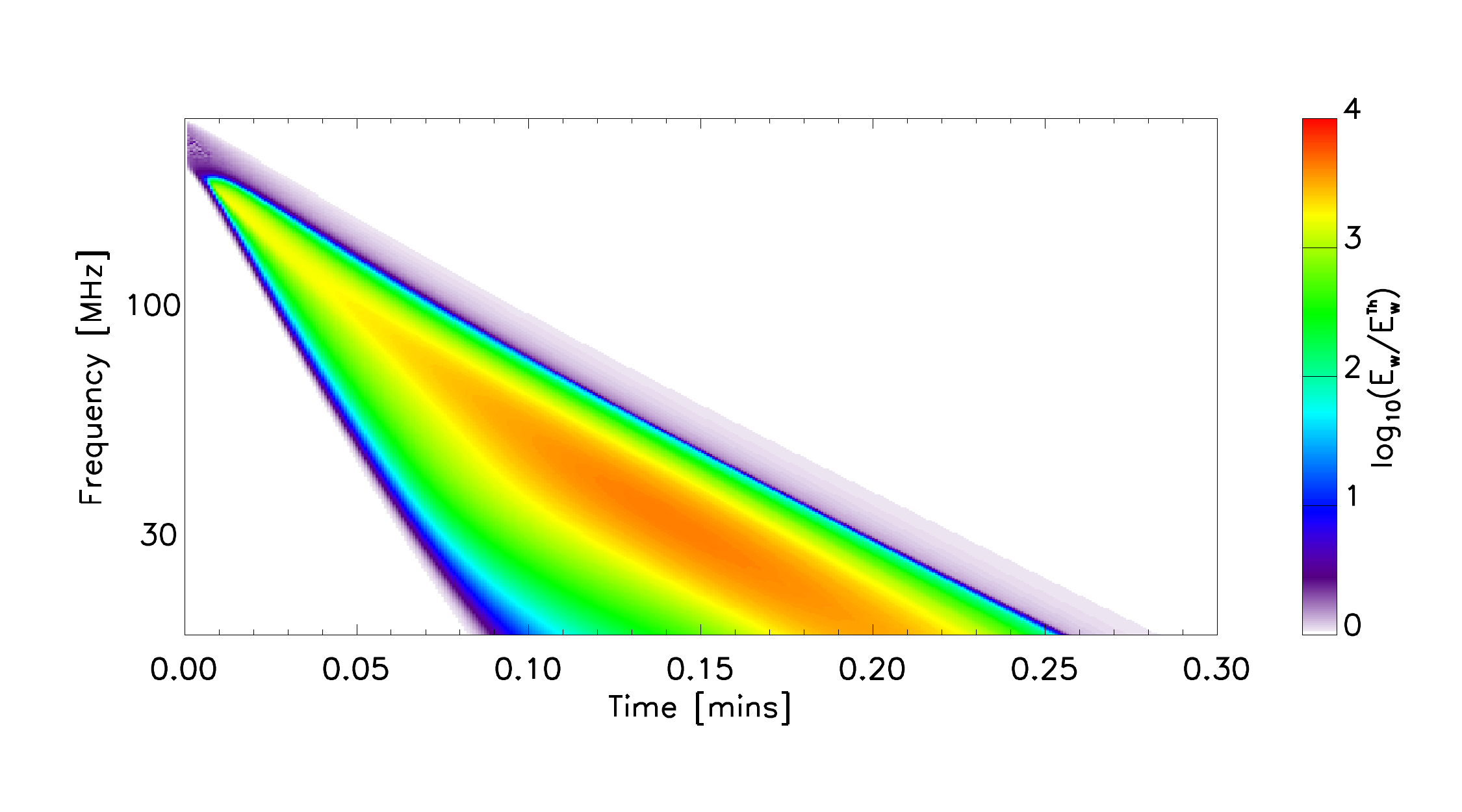}
  \includegraphics[width=0.47\textwidth,trim=22 20 85 35,clip]{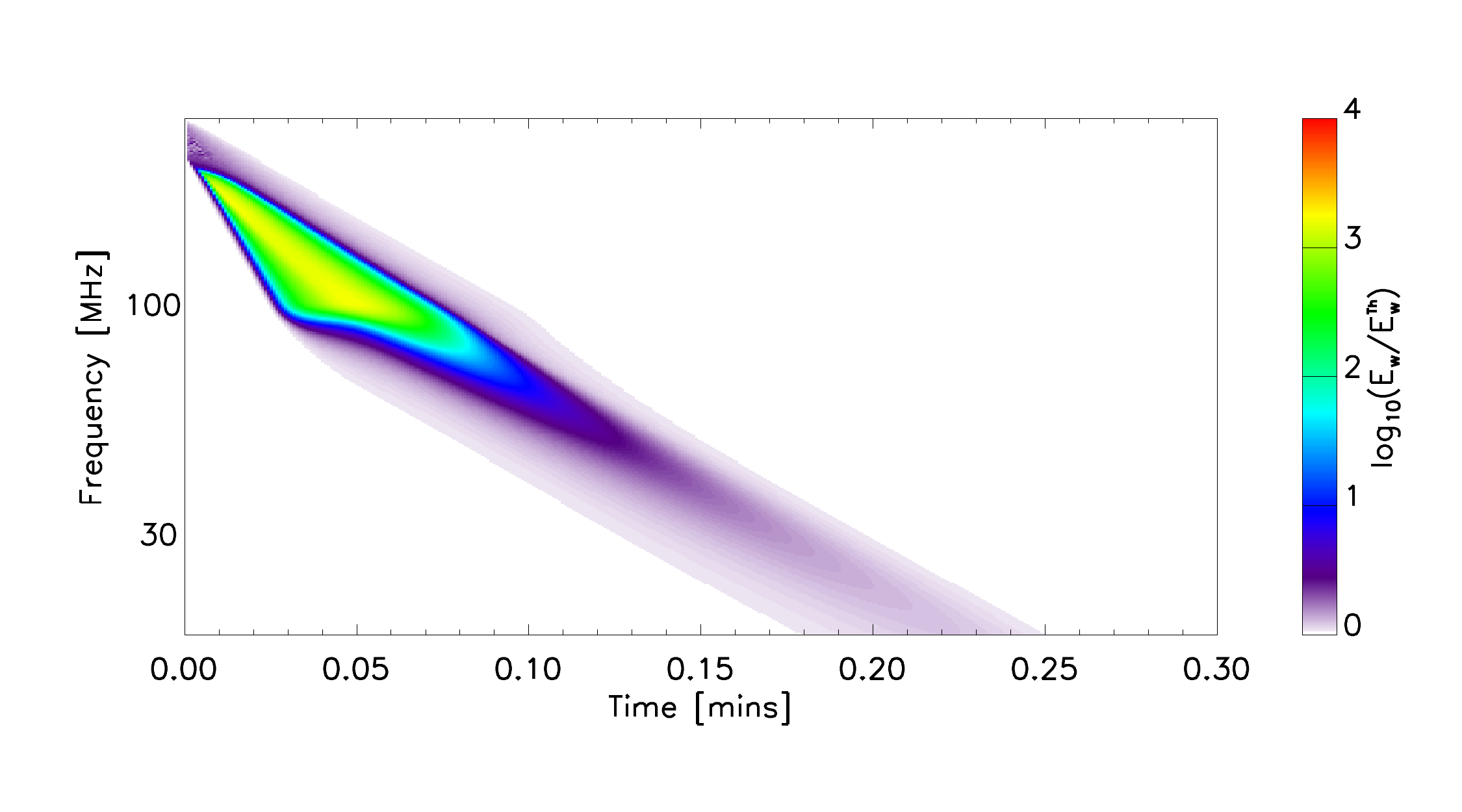}
  \includegraphics[width=0.52\textwidth,trim=22 20 25 35,clip]{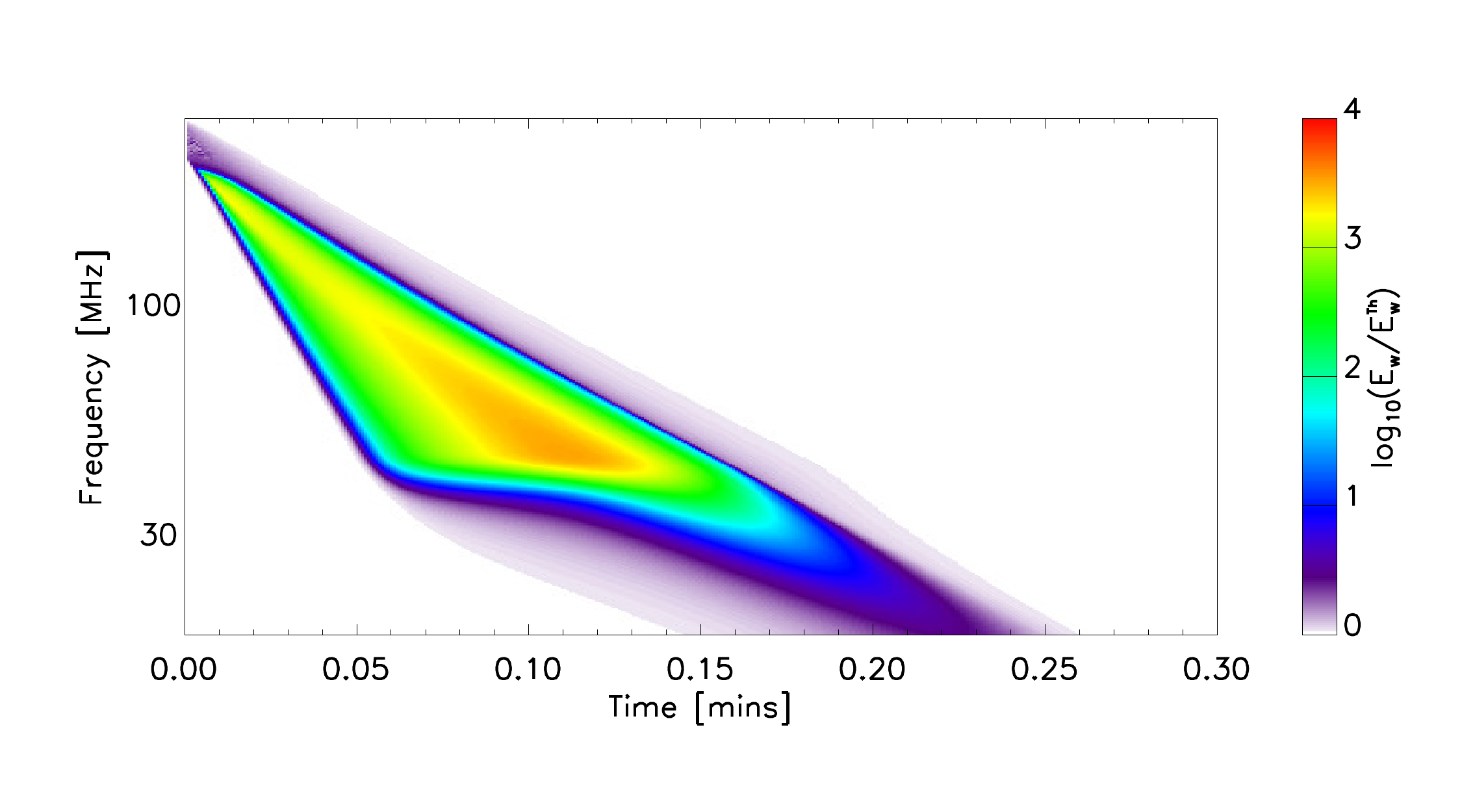}
  \includegraphics[width=0.47\textwidth,trim=22 20 85 35,clip]{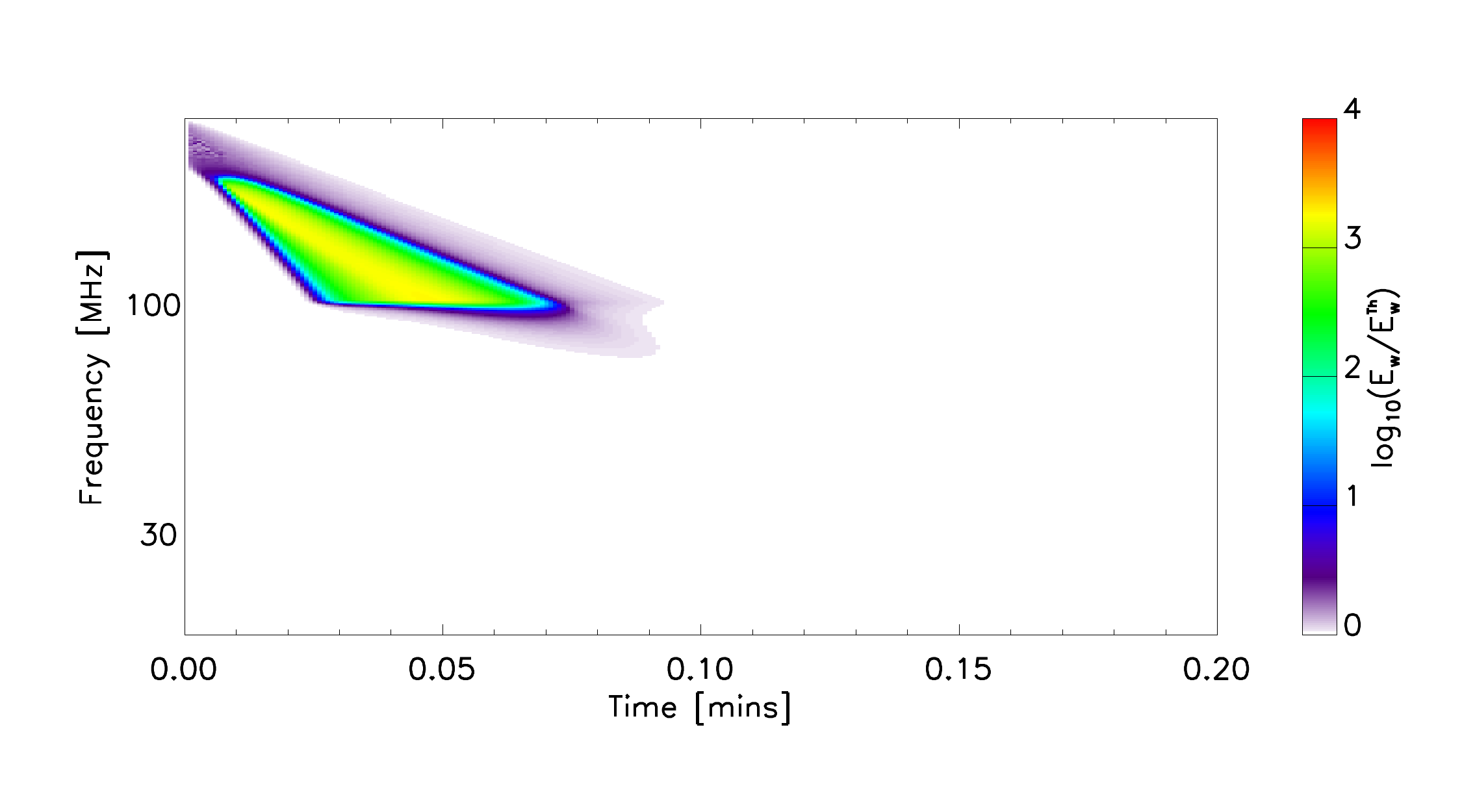}
  \includegraphics[width=0.52\textwidth,trim=22 20 25 35,clip]{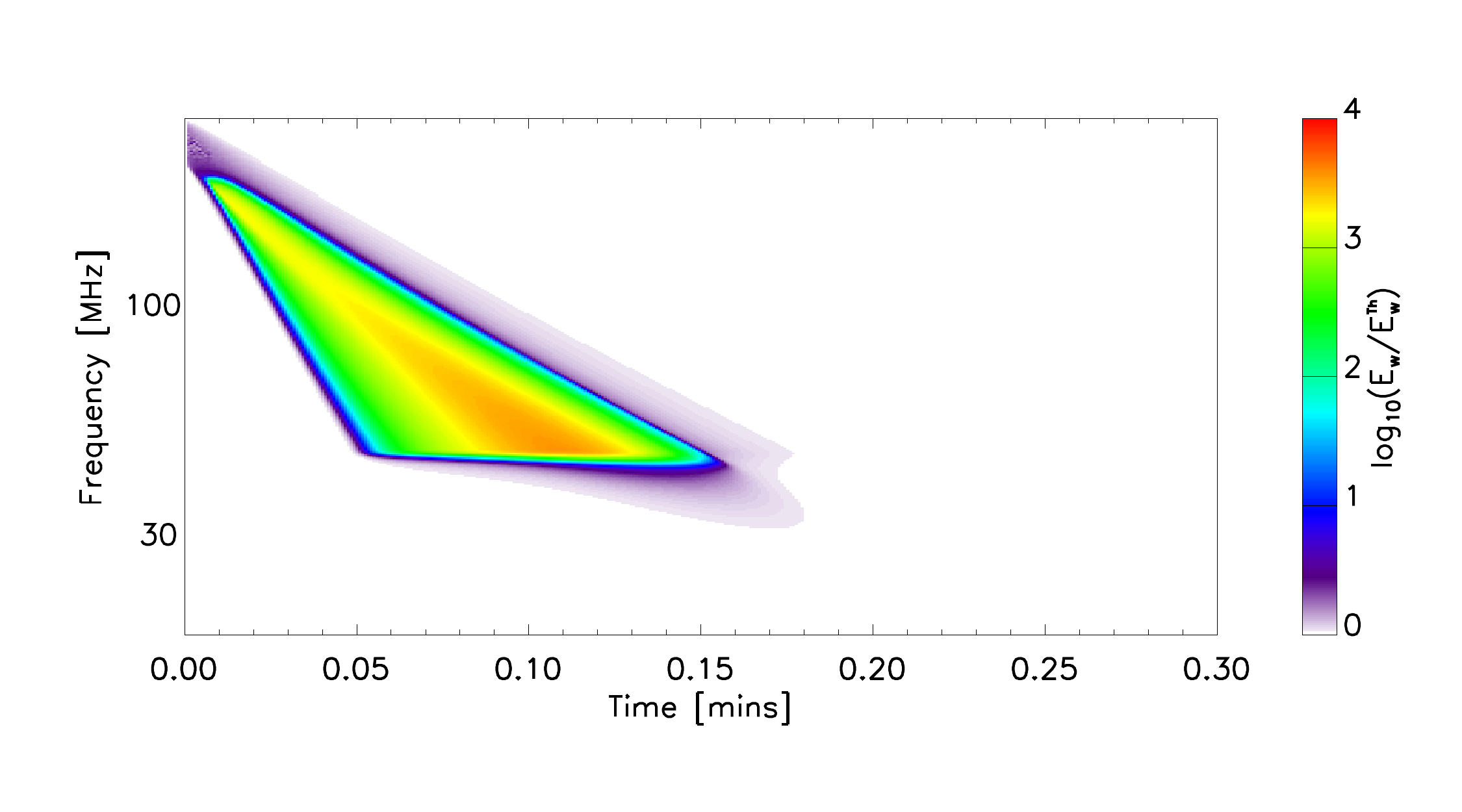}
\caption{Langmuir wave energy density $E_w /E_w^{init}$ in plasma frequency - time plane. Langmuir waves produced by the electron beam with parameters as in Figures \ref{fig:radial1}, \ref{fig:radial2} but with a magnetic flux tube that undergoes increased expansion at a critical height (\ref{eqn:double_expansion}).  Left panels:  The critical height is $r_c=2\times 10^{10}$~cm.  The severity of the extra expansion increases from top to bottom.   Right panels:  The critical height is $r_c=4\times 10^{10}$~cm.  The severity of the extra expansion increases from top to bottom as $Z_c=10^1, 10^3, 10^5$. }
\label{fig:expansion2}
\end{figure*}

In Figure \ref{fig:expansion2}, we set $\beta=3$ for the magnetic flux tube expansion, so that magnetic field decreases as $r^{-3}$.  To simulate a reduced expansion in the corona we increased the value of $r_0=10^{10}$~cm.  We set $r_c=2\times10^{10}$~cm and $r_c=4\times10^{10}$~cm to demonstrate expansion at two different heights above the photosphere that correspond roughly to 100~MHz and 50~MHz.  For each height we have varied the constant $Z_c=10^1, 10^3, 10^5$ to show different rates of expansion.

Figure \ref{fig:expansion2} shows the Langmuir wave energy density for the six different cases described above.   Langmuir wave energy density is clearly reduced after the electron beam reaches $r_c$ when the expansion is significant.  The two different frequencies of 100 MHz corresponding to $r_c=2\times 10^{10}~\rm{cm}$ and 40 MHz corresponding to $r_c=4\times 10^{10}~\rm{cm}$ are evident in Figure \ref{fig:expansion2} when $Z_c=10^5$, whereas Langmuir waves continue to be induced after the electron beam reaches $r_c$ when $Z_c=10$ (Figure \ref{fig:expansion2}).

\section{Density fluctuations in the background plasma}\label{sec:inhom}

The expansion of the guiding magnetic flux tubes is not the only factor that determines the stopping distance of a type III burst.  Density fluctuations in the background plasma are known to decrease the level of Langmuir waves induced by an electron beam \citep[e.g.][]{Ryutov1969,SmithSime1979,Melrose1980,Kontar2001,ReidKontar2010,Li_etal2012,Ratcliffe_etal2012,ReidKontar2013}.  Langmuir waves are shifted out of resonance with the electron beam by the density fluctuations that results in a decreased level of Langmuir wave turbulence.

The power spectrum of density fluctuations near the Earth has been observed to obey a Kolmogorov-type power law with a spectral index of $-5/3$ \citep[e.g.][]{Celnikier_etal1983,Celnikier_etal1987,Chen_etal2013c}.
Following \citet{ReidKontar2010}, the spectrum of density fluctuations was modelled with spectral index $-5/3$ between the wavelengths of $10^7$ and $10^{10}$~cm, so that the new perturbed density profile is given by the following equation
\begin{equation}\label{fluc}
n_e(r) = n_0(r)\left[1 + C(r)\sum_{n=1}^N\lambda_n^{\mu/2}\sin(2\pi r/\lambda_n + \phi_n)\right]\,,
\end{equation}
 where $N=1000$ is the number of perturbations, $n_0(r)$ is the initial unperturbed density (defined in Section \ref{sec:plasma}), $\lambda_n$ is the wavelength of $n$-th fluctuation, $\mu=5/3$ is the power-law spectral index in the power spectrum, and $\phi_n$ is the random phase of the individual fluctuations.  $C(r)$ is the normalisation that defines the level of density fluctuations
\begin{equation}
C(r) = \sqrt{\frac{2\langle \Delta n(r)^2 \rangle}{\langle n(r) \rangle^2\sum_{n=1}^N\lambda_n^{\mu}}}
\end{equation}
% where $\langle n(r) \rangle$ denotes the mean density at one point in space.
where the r.m.s. deviation of the density $\sqrt{\langle \Delta n(r)^2 \rangle}$ is taken so that near the Earth at $r=1$~AU  $\sqrt{\frac{\langle \Delta n(r=1 AU)^2 \rangle}{\langle n(r=1 AU) \rangle^2}}=0.1$.   \citet{ReidKontar2010} used numerical simulations of electron beams travelling to the Earth to estimate $C(r)$.  The best fit was a level of fluctuations that decreased as a function of distance using the formula
\begin{equation}\label{eqn:fluc_rad}
\frac{\Delta n}{n} = \sqrt{\frac{\langle \Delta n(r)^2 \rangle}{\langle n(r) \rangle^2}} = \left(\frac{n_0(1 AU)}{n_0(r)}\right)^{\Psi} \sqrt{\frac{\langle \Delta n(r=1 AU)^2 \rangle}{\langle n(r=1 AU) \rangle^2}}
\end{equation}
where $\Psi=0.25$.  This results in a level of fluctuations (denoted now for simplicity as $\Delta n/n$) at the Sun that is roughly $1\%$ of the level at the Earth, or $\Delta n/n=10^{-3}$.  The value increases as a function of distance to $\Delta n/n=10^{-1}$ at the Earth.

\subsection{Simulations}

\begin{figure*}\center
  \includegraphics[width=0.47\textwidth,trim=22 20 85 35,clip]{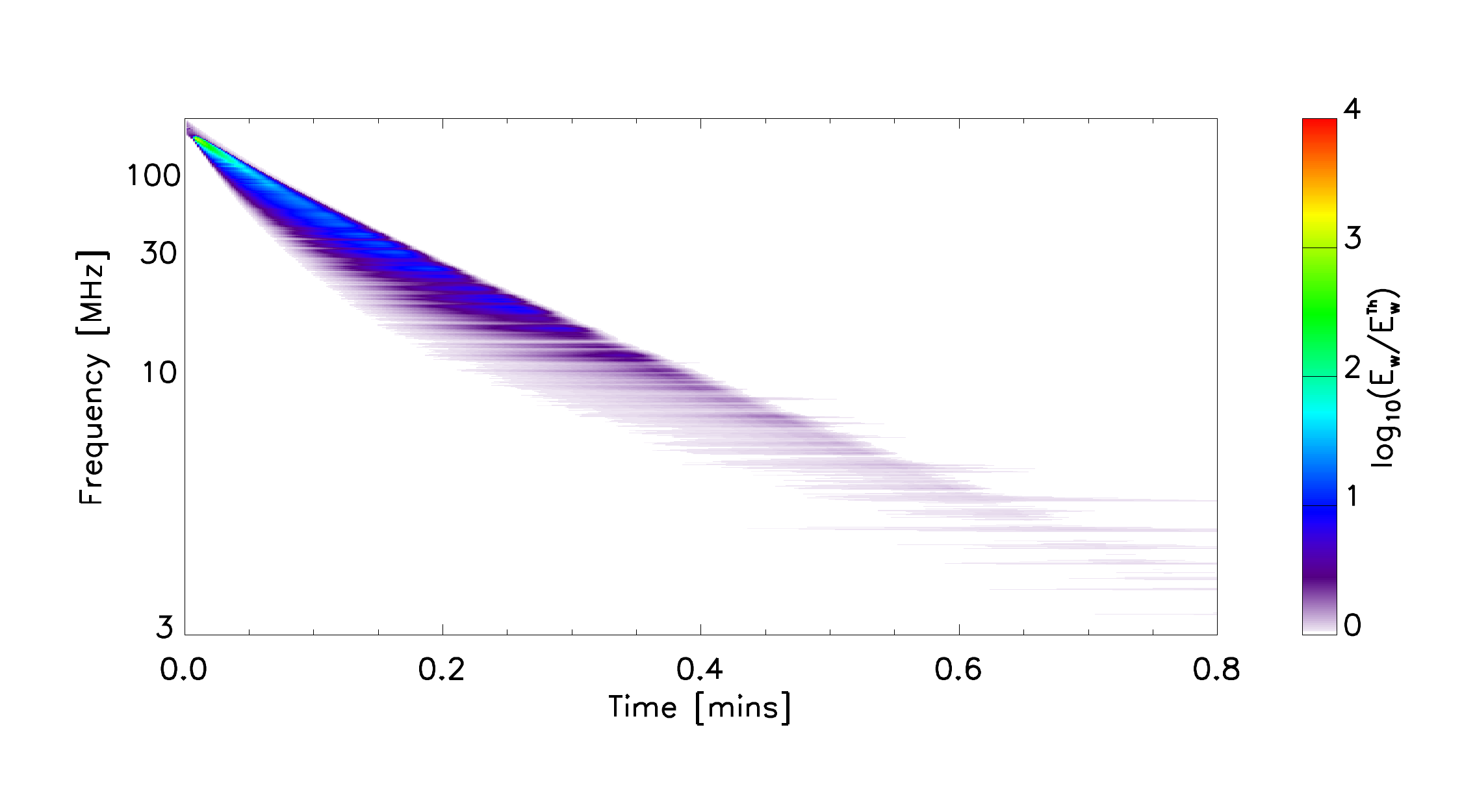}
  \includegraphics[width=0.52\textwidth,trim=22 20 25 35,clip]{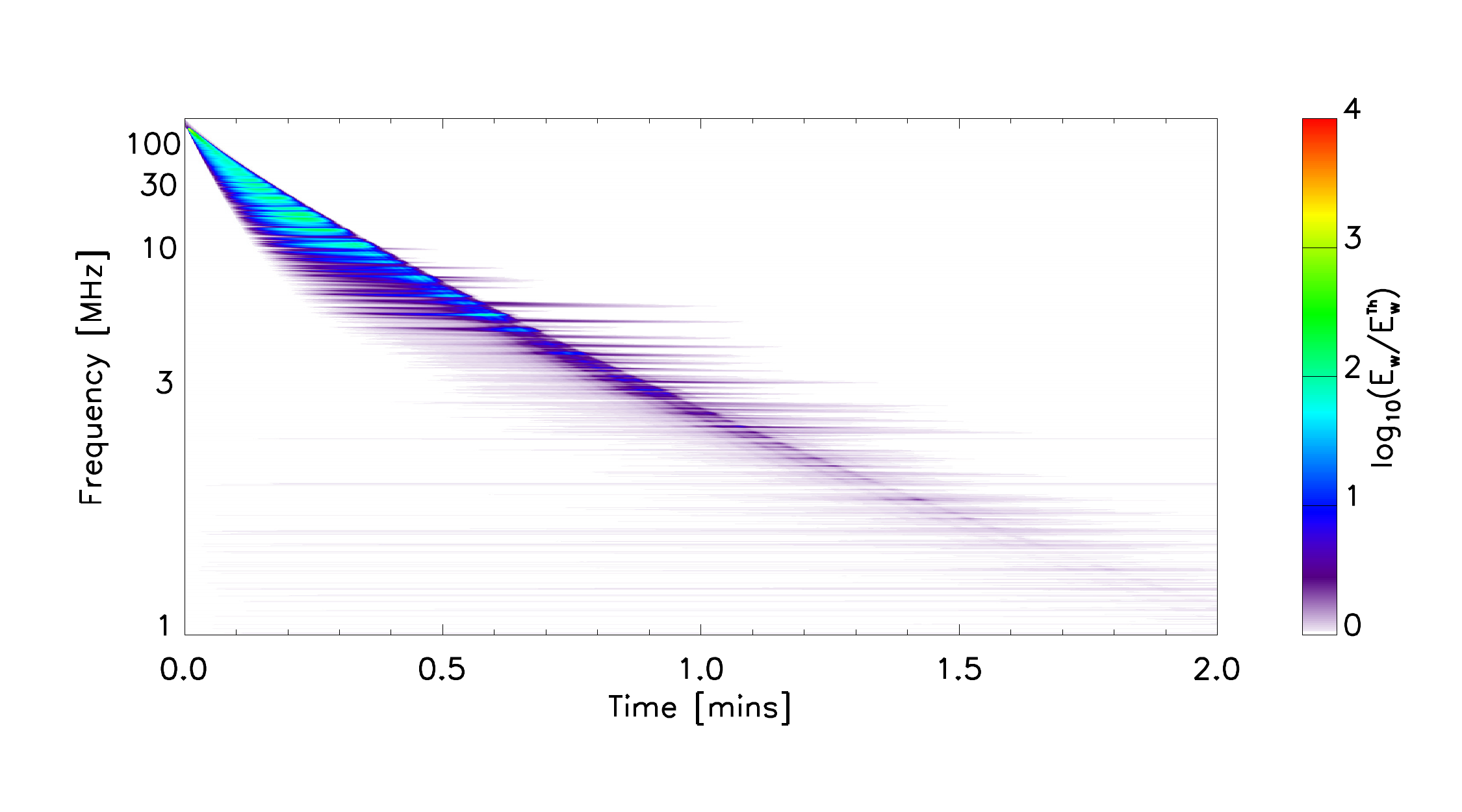}
  \includegraphics[width=0.47\textwidth,trim=22 20 85 35,clip]{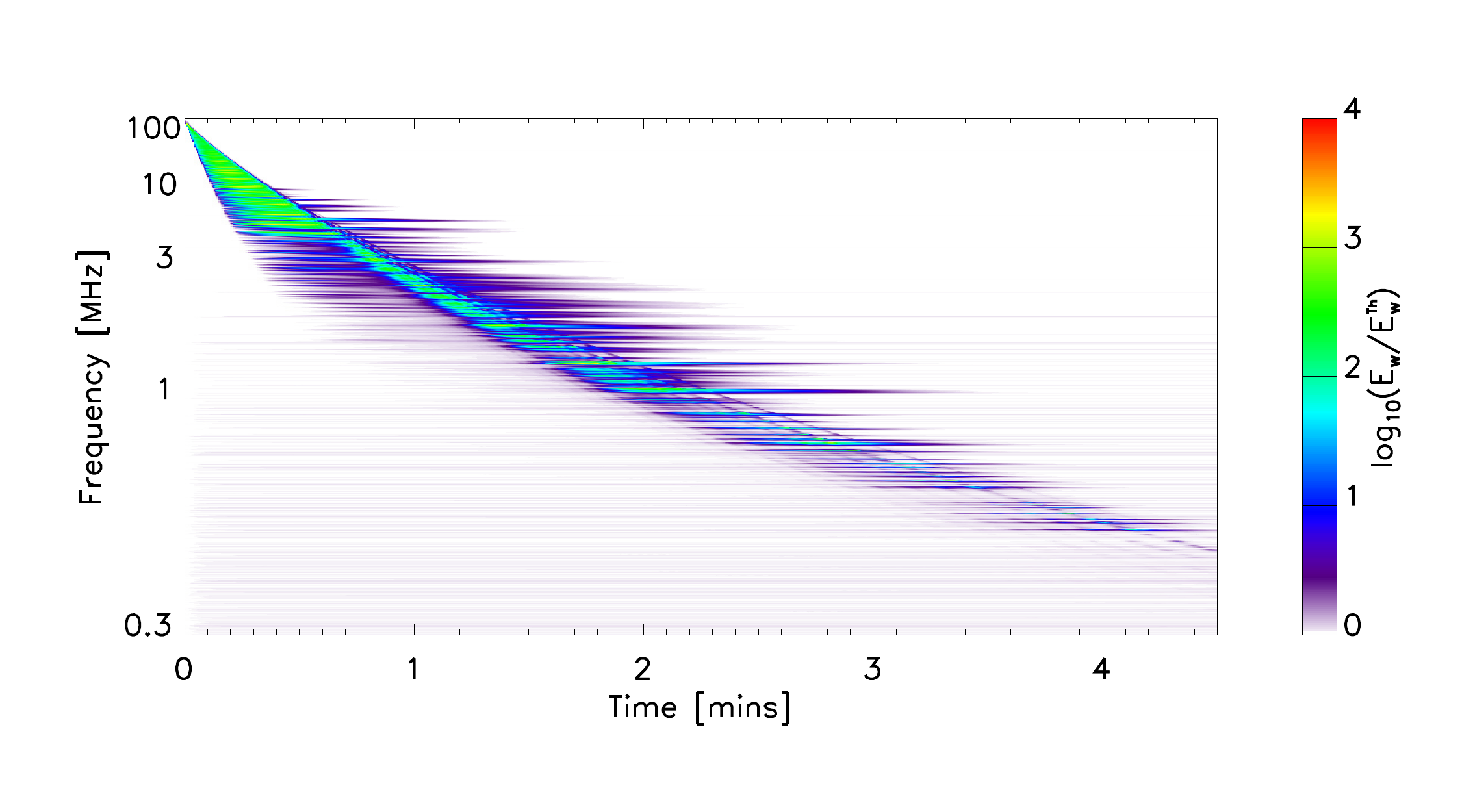}
  \includegraphics[width=0.52\textwidth,trim=22 20 25 35,clip]{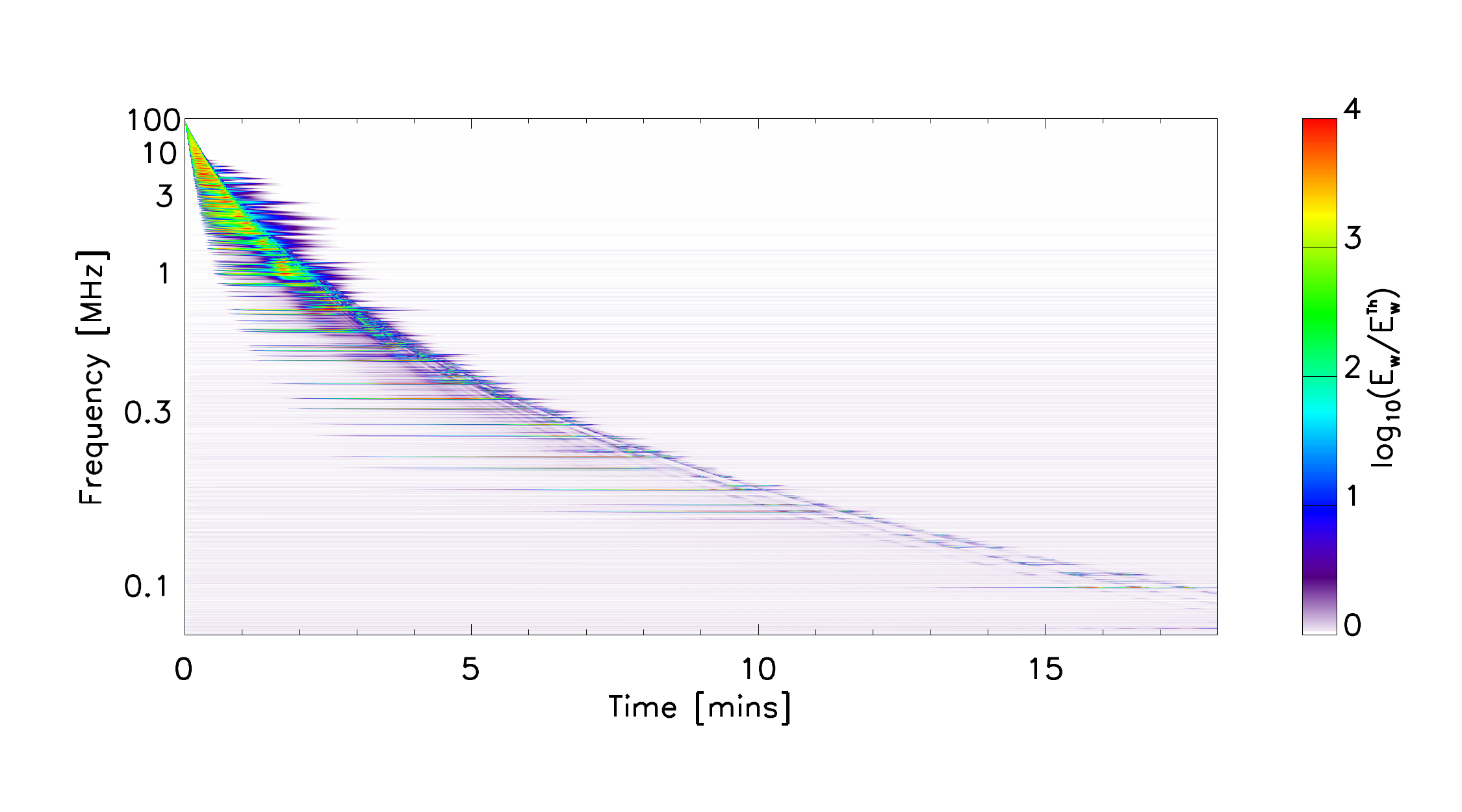}
\caption{Langmuir wave energy density $E_w /E_w^{init}$ in the plasma frequency - time plane. The electron beam initial parameters are as in Figures \ref{fig:radial1}, \ref{fig:radial2} except that the background plasma includes density fluctuations.  Again, the coefficient that models the magnetic flux tube expansion is $\beta=3.5$ (top left), $\beta=3.0$ (top right), $\beta=2.5$ (bottom left), $\beta=2.0$ (bottom right).  Note the different values of the frequency and time axis between each graph.}
\label{fig:inhom1}
\end{figure*}

Using the same beam and plasma parameters as Section \ref{sec:normal} we conducted the simulations with the inclusion of a fluctuating background plasma. We varied the expansion of the magnetic flux tubes using $\beta=2,~2.5,~3.0,~3.5$.  The normalised spectral energy density of Langmuir waves $E_w^{max}/E_w^{init}$ are presented as a function of time and plasma frequency in Figure \ref{fig:inhom1}.  The bursty nature of the Langmuir wave energy density is evident, especially comparing Figure \ref{fig:inhom1} to Figure \ref{fig:radial1}. The bursty nature of the Langmuir energy density in Figure \ref{fig:inhom1} is increased for lower frequencies because of the model that we used for density fluctuations.  Equation (\ref{eqn:fluc_rad}) has a larger intensity of density fluctuations farther away from the Sun.  

Figure \ref{fig:radial2} (right) shows how the different levels of magnetic field expansion affects the minimum frequency of the background plasma that certain levels of Langmuir wave energy density are induced.  A comparison with the case without fluctuations reveals that in general the inclusion of density fluctuations increases this minimum frequency.  For instance at $\beta=2.0$ an increased Langmuir wave energy density was not observed below 0.08~MHz compared to the 0.02~MHz with density fluctuations were not included.  Density fluctuations in the background plasma are therefore another important parameter in the background plasma that affects the stopping frequency of type III bursts.  Increasing the r.m.s. deviation of the density $\Delta n/n$ given in Equation (\ref{eqn:fluc_rad}) increases the suppression of Langmuir waves induced by the electron beam density, and vice versa.

We also point out that for any single value of $\beta$, the points in Figure \ref{fig:radial2} are more closely spaced when you consider density fluctuations in the background plasma. Indeed, all points are almost on top of each other when $\beta=2.0$ or $\beta=2.5$. These points are due to the spikes in the spatial distribution of Langmuir wave energy that reached at least $10^3$ times that of the background level.  The spikes are able to produce higher Langmuir wave energy density at lower background frequencies.  For example a level around $10^3$ is only observed at 5~MHz without fluctuations but at 0.5~MHz with fluctuations.  However, the total Langmuir wave energy is less when density fluctuations are considered.

åç% \textbf{It is also worth noting that the presence of enhanced density fluctuations can increase the nonlinear conversion process of radio emission from Langmuir waves resulting in a lower energy density of Langmuir waves being required for radio emission.  For the simulations shown in Figure \ref{fig:radial2} there is very little Langmuir wave energy density below the stopping frequencies found (observable in Figure \ref{fig:inhom1}.  Given the order of magnitude difference between the stopping frequencies observed in the two graphs in Figure \ref{fig:radial2} we still conclude that the presence (and intensity) of density fluctuations will increase the stopping frequency of type III bursts.}

\section{Role of initial electron beam parameters} \label{sec:beam_params}

In this section we explore how the different initial electron beam parameters can affect the stopping frequency of type III bursts.  Specifically we look at the number of injected electrons and how they are distributed in energy space.

\subsection{Injected beam density} \label{sec:density}

Obviously, the number of electrons we inject into the simulations is important when determining the level of Langmuir waves that is induced and consequently at what frequency the beam stops wave generation.
As described in Section \ref{sec:beam_params}, the simulation parameter that dictates the number density of electrons injected into the simulation is $n_{beam}$\footnote{$n_{beam}$ is the time integrated beam density injected at the centre of the acceleration region.  However, given the fast injection characteristic time of $10^{-3}$~s, we refer to it simply as the beam density.}.  Our simulations are set up as described in Section \ref{sec:model_setup} with the guiding flux tube characterised using $\beta=2.5$.  We have varied the beam density over one order of magnitude such that $n_{beam} = 5\times 10^6~\rm{cm}^{-3},~2\times 10^6~\rm{cm}^{-3},~1\times 10^6~\rm{cm}^{-3},~5\times 10^5~\rm{cm}^{-3}$.

\begin{figure*}
  \includegraphics[width=0.49\textwidth]{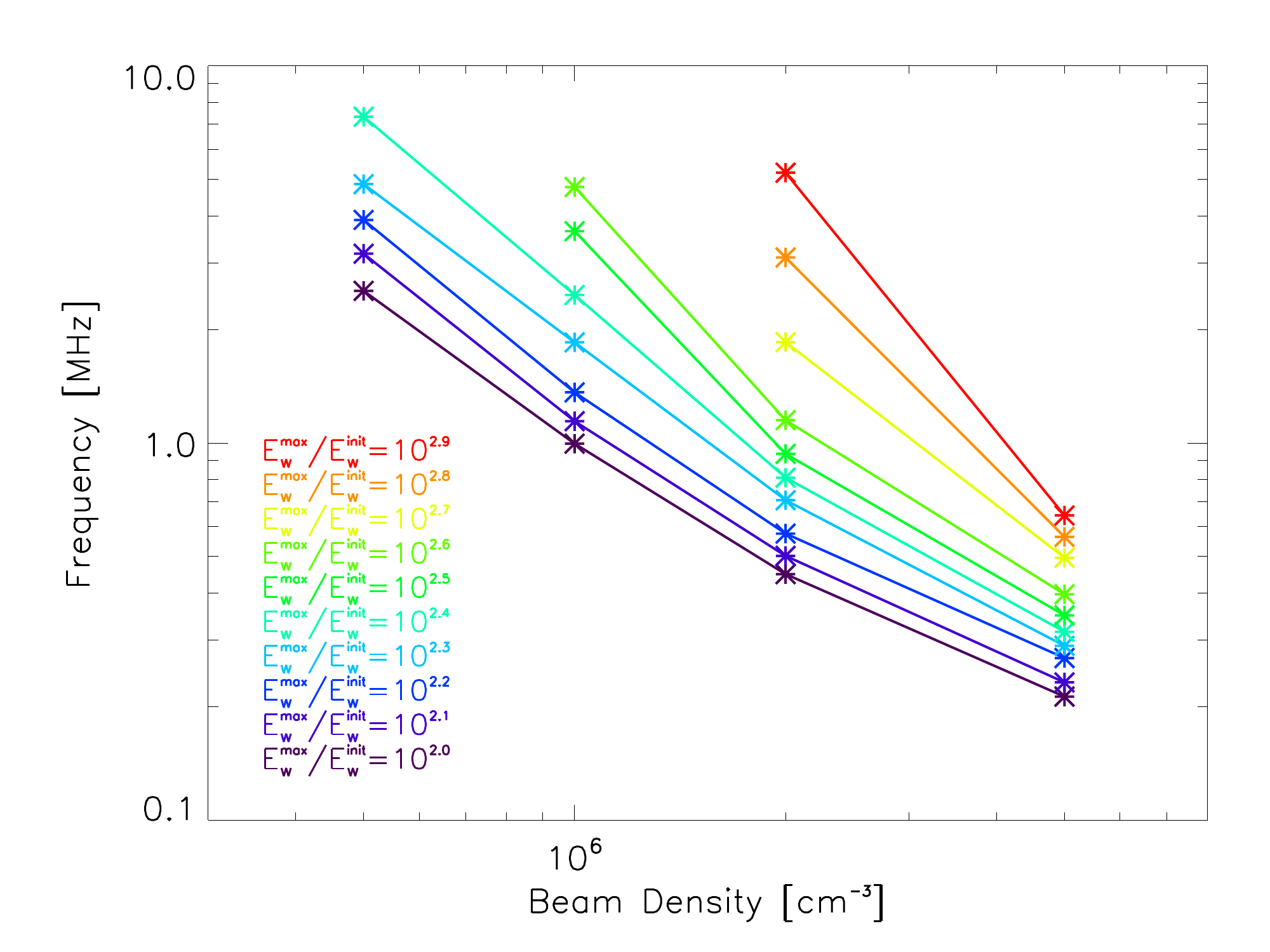}
  \includegraphics[width=0.49\textwidth]{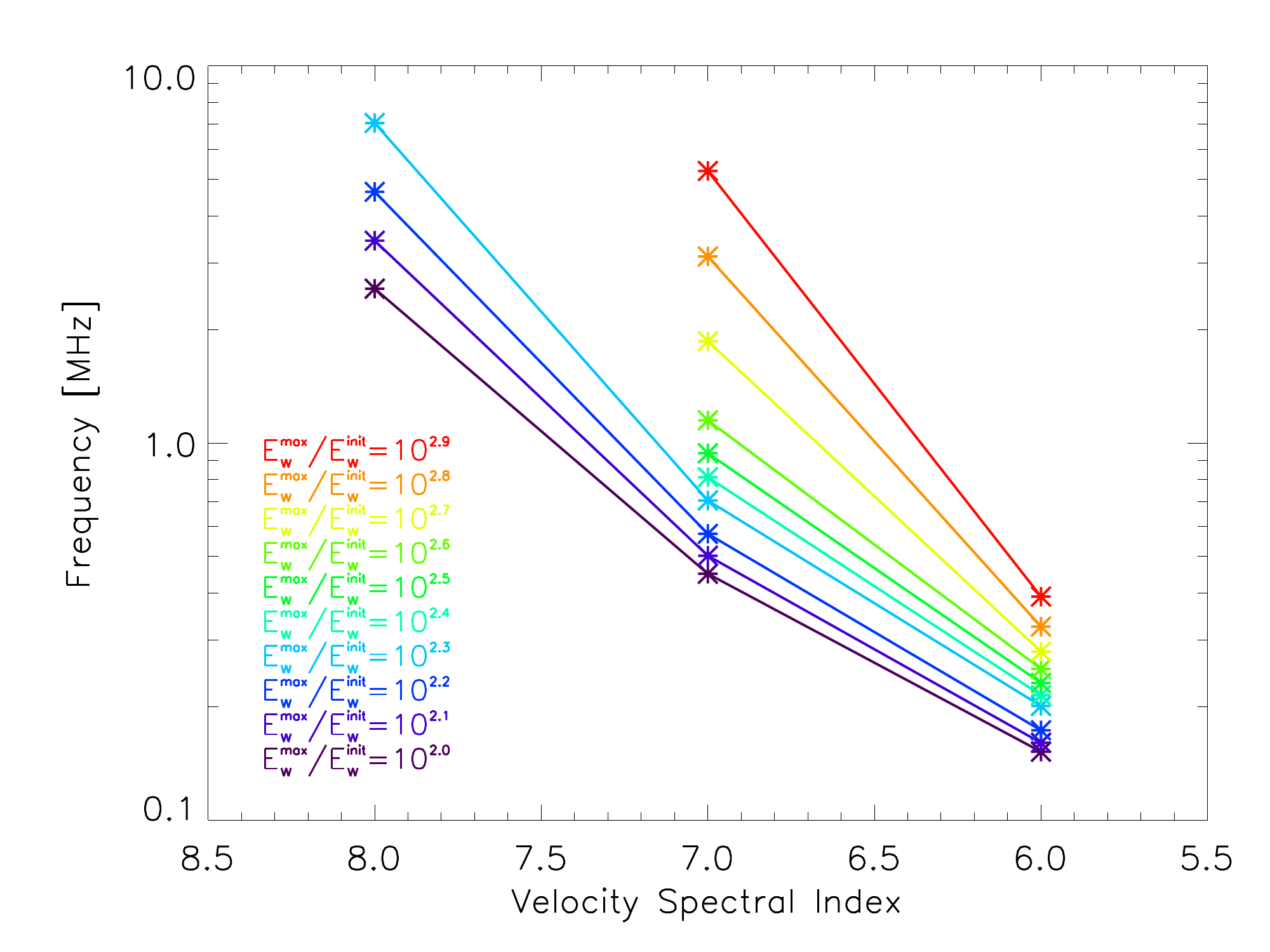}
\caption{The minimum frequency (stopping frequency) where an electron beam is able to induce certain amount of Langmuir wave energy density (see color-coded lines).  Left: plotted against beam density.  Right: plotted against velocity spectral index.  The different coloured lines represent different levels of Langmuir wave energy (normalised by the thermal level) between $10^2$ and $10^3$.}
\label{fig:dens_si}
\end{figure*}

 Similar to the previous sections, we have plotted the normalised maximum Langmuir wave energy density $E_w^{max}/E_w^{init}$ as a function of frequency in Figure \ref{fig:dens_si}.  As expected, the beams with higher initial densities are able to excite higher levels of Langmuir wave energy at lower plasma frequencies (or at distances farther away from the Sun).  As the initial number of electrons in the beam is reduced, the beam is less and less able to generate Langmuir waves farther out.  For beam density $5\times10^5~\rm{cm}^{-3}$ the beam is not able to excite $E_w^{max}/E_w^{init}>10^{2.4}$ other than when it initially became unstable at frequencies $>100$~MHz, similar to the beams that experience high magnetic flux tube expansion in Figure \ref{fig:expansion2}.

\subsection{Injected beam energy distribution} \label{sec:spectralindex}

The distribution of beam electrons as a function of energy is very important for the production of Langmuir waves.  Exactly what distribution we inject into the simulation will have a very significant effect on the minimum frequency the beam can interact with Langmuir waves.  The energy distribution is governed by the spectral index $\delta$ defined in Equation (\ref{eqn:velocity}).  We demonstrate the importance by using simulations with the same parameters as Section \ref{sec:normal}, keeping the magnetic field expansion static using $\beta=2.5$, and varying the velocity spectral index $\delta=6,7,8$.

We plot the normalised maximum Langmuir wave energy density $E_w^{max}(r)/E_w^{init}(r)$ for these three initial spectral indices as a function of frequency in Figure \ref{fig:dens_si}.  We can see the frequencies corresponding to a high peak level of Langmuir wave energy density dramatically decreases as we decrease the initial velocity spectral index.  As we have normalised the number density of electrons injected, if we lower the spectral index, more high energies are injected.  There are two reasons this leads to higher level of Langmuir wave turbulence:
\begin{itemize}

\item The injection altitude is relatively low in the corona ($f_{pe}=500$~MHz).  The background density is $2\times10^9~\rm{cm}^{-3}$ where the collision rate is not negligible.  Therefore the low energy electrons (e.g. $\lesssim 6$~keV) collisionally lose energy before the beam escapes to a lower density background plasma.  As demonstrated in \citet{ReidKontar2013} the electrons with energy $1.5-6$~keV will lose 99\% of their energy after only 0.1~R$_{\odot}$ of travel.  A lower spectral index (harder spectrum) means that the beam loses less energy in the low corona and retains more energy to transfer to Langmuir waves. 
    
\item Since the growth rate of Langmuir waves at one point in space is proportional to $v^2 \partial f/\partial v$, the $v^2$ term means that a flatter initial electron spectrum will increase it.

\end{itemize}

\section{Discussion and conclusions} \label{sec:conclusion}

To analyse the stopping frequency of type III radio bursts the outward propagation of deka-keV electrons have been modelled from a flaring acceleration site through the solar corona and inner heliosphere.  The wave-particle interaction between electrons and Langmuir waves, Coulomb collisions, Langmuir wave refraction, and the radial expansion of the guiding magnetic flux tubes have been included.  We have investigated the lowest frequency that the beam causes enhanced levels of Langmuir waves, and hence radio waves, for different electron beam and background plasma properties.  

The radial expansion of the guiding magnetic flux tube significantly affects the distances from the Sun that electrons can induce high levels of Langmuir wave turbulence (and hence characteristic type III burst frequencies). The electron beam follows the expanding magnetic flux tube and consequently decreases in density as a function of distance from the Sun.  The decrease in density eventually quenches the instability that causes Langmuir wave generation. This scenario could be the cause of type III bursts that are observed only at high frequencies although it is not the only mechanism that can produce such type III bursts.

Background density fluctuations caused an electron beam to cease producing Langmuir waves at shorter distances from the Sun. However, their effect on stopping frequencies was not as significant as varying the rate of magnetic flux tube expansion.  Due to the clumping of Langmuir waves in space, caused by the density fluctuations, we were able to observe high level of Langmuir waves at reasonably low frequencies but the overall level of Langmuir waves was reduced.

Varying the properties of the injected electron beam also affects the lowest frequency where Langmuir waves production is significant.  Decreasing the density of the injected electron beam or increasing the spectral index of the injected electron beam reduces the distance the electron beam can travel before it stops producing high levels of Langmuir wave turbulence.  We then expect that less dense electron beams or beams with high spectral indices will have higher type III stopping frequencies.  It is possible that injection of electrons at high altitudes $\gtrsim 0.1~\rm{R}_{\odot}$ would be less sensitive to the spectral index of the beam in respect to the type III stopping frequency due to a reduced role of Coulomb collisions.

How do these finding relate to some of the hypothesised reasons in \citet{Leblanc_etal1995}?  We find that all the different effects that govern the number of high energy electrons play some roles in the type III stopping frequency: the expansion of the guiding magnetic flux tubes, the injected beam density and the injected spectral index.  We also found that large scale density fluctuations in the background electron density play a role, although a high level of density fluctuations may still cause irregular radio emission caused by the localised clumping of Langmuir waves. The hypothesis that we have not checked is how the other non-linear plasma processes required to convert Langmuir waves into radio waves affect the stopping frequency of type III bursts.  What we have also explained is the prevalence for stronger type III bursts to have lower stopping frequencies, a result observed by \citet{Leblanc_etal1995,Leblanc_etal1996,Dulk_etal1996}.  The beams with more deca-keV electrons produce in general higher levels of Langmuir waves, and consequently make brighter type IIIs and have lower stopping frequencies.  The upcoming missions Solar Orbiter and Solar Probe Plus should help to observationally study the radial plasma properties with in-situ measurements.

\begin{figure}\centering
  \includegraphics[width=0.49\textwidth,trim=22 20 25 35,clip]{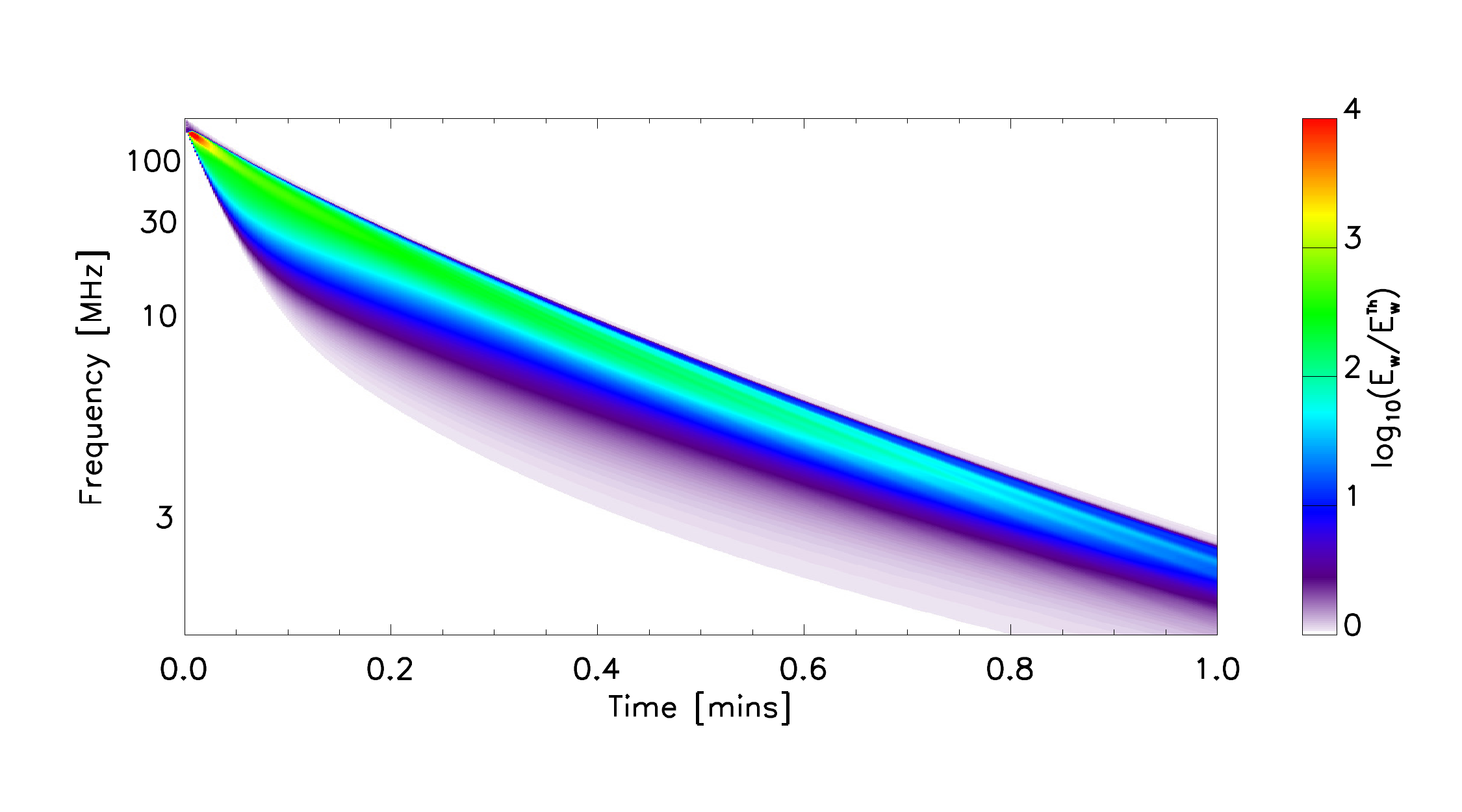}
\caption{Langmuir wave energy density $E_w/E_{init}$ in plasma 
frequency - time plane. The coefficient that models the magnetic flux tube expansion is $\beta=3.5$ and the velocity spectral index of injected electrons is $\alpha=5$.}
\label{fig:silow}
\end{figure}

Magnetic flux tubes have been observed to expand super-radially (rates faster than $r^2$) in the corona with type III radio emission following the magnetic flux tube expansion \citep[e.g.][]{Klein_etal2008}.  Whilst this expansion can be large, it is very difficult to have a large expansion such that $B\propto r^{-3.5}$ and produce type III radio emission.  The electron beam decreases in density too fast.  To demonstrate this we injected an electron beam with the same characteristics described in Section \ref{sec:beam} but with a spectral index of $5$ in velocity space.  Such a low spectral index is on the verge of what is inferred observationally from hard X-rays \citep[see e.g.][]{2005SoPh..232...63K}. Figure \ref{fig:silow} shows how the energy density of Langmuir waves varies as a function of time and frequency.  We only observe a high level of Langmuir wave energy at the high frequencies and not at the low frequencies.  A flare with such a low spectral index would almost certainly produce type III radio emission below $0.1$~MHz, similar to Figure \ref{fig:obs1}.  We conclude that we would not observe type III bursts if the solar magnetic field ever expands at such a super-radial rate.

From type III burst simulations \citep{Ratcliffe_etal2014}, it is evident that the spectrum of Langmuir wave energy density does not exactly reflect the spectrum of radio waves that are observed.  The requirement for backscattered Langmuir waves to interact with forward propagating Langmuir waves in the production of harmonic emission means that spectral characteristics of Langmuir waves are important, not just the absolute energy density.  However, the radial expansion of the magnetic flux tube reduces the Langmuir wave production. Therefore our conclusions on stopping frequency are expected to be mirrored in escaping radio emission, although a detailed study including the generation of radio emission can explore additional effects of wave-wave plasma processes on the type III stopping frequencies.

\begin{acknowledgements}
Hamish Reid acknowledges funding from a SUPA Advanced Fellowship.  Eduard Kontar acknowledges funding from an STFC consolidated grant.  Support from a Marie Curie International Research Staff Exchange Scheme Radiosun PEOPLE-2011-IRSES-295272 RadioSun project is greatly appreciated.  This work benefited from the Royal Society grant RG130642.
\end{acknowledgements}
 
\bibliographystyle{aa}
\bibliography{radial3_arxiv}

\end{document}